\documentclass[aps, prx, showpacs,reprint, amsmath, amssymb, superscriptaddress, floatfix]{revtex4-2}
\usepackage{graphicx}
\usepackage{physics}
\usepackage{txfonts}
\usepackage{lmodern}
\usepackage{mathtools}
\usepackage{multirow}
\begin{document}
\title{A mid-circuit erasure check on a dual-rail cavity qubit using the \\ joint-photon number-splitting regime of circuit QED}
\author{Stijn J. de Graaf}
\thanks{These authors contributed equally to this work.\\ stijn.degraaf@yale.edu \\ sophia.xue@yale.edu}
\affiliation{Department of Applied Physics, Yale University, New Haven, Connecticut 06511, USA}
\affiliation{Yale Quantum Institute, Yale University, New Haven, Connecticut 06511, USA}
\author{Sophia H. Xue}
\thanks{These authors contributed equally to this work.\\ stijn.degraaf@yale.edu \\ sophia.xue@yale.edu}
\affiliation{Department of Applied Physics, Yale University, New Haven, Connecticut 06511, USA}
\affiliation{Yale Quantum Institute, Yale University, New Haven, Connecticut 06511, USA}

\author{Benjamin J. Chapman}
\affiliation{Department of Applied Physics, Yale University, New Haven, Connecticut 06511, USA}
\affiliation{Yale Quantum Institute, Yale University, New Haven, Connecticut 06511, USA}

\author{James D. Teoh}
\affiliation{Yale Quantum Institute, Yale University, New Haven, Connecticut 06511, USA}
\affiliation{Department of Physics, Yale University, New Haven, Connecticut 06511, USA}
\affiliation{Quantum Circuits, Inc., 25 Science Park, New Haven, Connecticut 06511, USA}

\author{Takahiro Tsunoda}
\affiliation{Department of Applied Physics, Yale University, New Haven, Connecticut 06511, USA}
\affiliation{Yale Quantum Institute, Yale University, New Haven, Connecticut 06511, USA}

\author{Patrick Winkel}
\affiliation{Department of Applied Physics, Yale University, New Haven, Connecticut 06511, USA}
\affiliation{Yale Quantum Institute, Yale University, New Haven, Connecticut 06511, USA}

\author{John W. O. Garmon}
\affiliation{Yale Quantum Institute, Yale University, New Haven, Connecticut 06511, USA}
\affiliation{Department of Physics, Yale University, New Haven, Connecticut 06511, USA}

\author{Kathleen M. Chang}
\affiliation{Department of Applied Physics, Yale University, New Haven, Connecticut 06511, USA}
\affiliation{Yale Quantum Institute, Yale University, New Haven, Connecticut 06511, USA}

\author{Luigi Frunzio}
\affiliation{Department of Applied Physics, Yale University, New Haven, Connecticut 06511, USA}
\affiliation{Yale Quantum Institute, Yale University, New Haven, Connecticut 06511, USA}
\affiliation{Quantum Circuits, Inc., 25 Science Park, New Haven, Connecticut 06511, USA}

\author{Shruti Puri}
\affiliation{Department of Applied Physics, Yale University, New Haven, Connecticut 06511, USA}
\affiliation{Yale Quantum Institute, Yale University, New Haven, Connecticut 06511, USA}

\author{Robert J. Schoelkopf}
\thanks{robert.schoelkopf@yale.edu}
\affiliation{Department of Applied Physics, Yale University, New Haven, Connecticut 06511, USA}
\affiliation{Yale Quantum Institute, Yale University, New Haven, Connecticut 06511, USA}
\affiliation{Department of Physics, Yale University, New Haven, Connecticut 06511, USA}
\affiliation{Quantum Circuits, Inc., 25 Science Park, New Haven, Connecticut 06511, USA}

\begin{abstract}

Quantum control of a linear oscillator using a static dispersive coupling to a nonlinear ancilla underpins a wide variety of experiments in circuit QED. Extending this control to more than one oscillator while minimizing the required connectivity to the ancilla would enable hardware-efficient multi-mode entanglement and measurements. We show that the spectrum of an ancilla statically coupled to a single mode can be made to depend on the joint photon number in two modes by applying a strong parametric beamsplitter coupling between them. This `joint-photon number-splitting' regime extends single-oscillator techniques to two-oscillator control, which we use to realize a hardware-efficient erasure check for a dual-rail qubit encoded in two superconducting cavities. By leveraging the beamsplitter coupling already required for single-qubit gates, this scheme permits minimal connectivity between circuit elements. Furthermore, the flexibility to choose the pulse shape allows us to limit the susceptibility to different error channels. We use this scheme to detect leakage errors with a missed erasure fraction of $(9.0 \pm 0.5)\times10^{-4}$, while incurring an erasure rate of $2.92 \pm 0.01\%$ and a Pauli error rate of $0.31 \pm 0.01\%$, both of which are dominated by cavity errors.

\end{abstract}
\maketitle
\section*{INTRODUCTION}
Controlling the state of an oscillator is a powerful resource, enabling the implementation of hardware efficient error-correcting codes for quantum computing \cite{ofek_extending_2016, fluhmann_encoding_2019, hu_quantum_2019, campagne-ibarcq_quantum_2020, gertler_protecting_2021, ni_beating_2023, sivak_real-time_2023}, simulations of bosonic systems \cite{braumuller_analog_2017, hu_simulation_2018, owens_quarter-flux_2018, wang_efficient_2020, wang_observation_2023} and the generation of metrologically useful states for quantum-enhanced sensing \cite{mccormick_quantum-enhanced_2019, wang_heisenberg-limited_2019, dixit_searching_2021}. In circuit quantum electrodynamics (cQED) \cite{wallraff2004strong,blais_cavity_2004}, where these oscillators take the form of standing modes in microwave resonators, most of the techniques developed for single-oscillator control \cite{heeres_cavity_2015, heeres_implementing_2017, eickbusch_fast_2022, diringer_conditional-not_2024, reinhold_error-corrected_2020, Rosenblum_fault-tolerant_2018, sun_tracking_2014, roy2024synthetic} rely on a static dispersive coupling between the oscillator and a nonlinear ancilla qubit \cite{schuster_resolving_2007}.

Moving beyond control of a single linear mode affords new capabilities, including the generation of multi-mode entanglement \cite{Wang-deterministic_2011, gao_entanglement_2019, chakram_multimode_2022} and measurements of joint properties of multiple modes \cite{Wang-cat-2016, wollack2022quantum, Gertler-realization_2023, koottandavida_erasure_2024}. In the context of quantum error correction, it both enables gates between qubits encoded in individual modes \cite{rosenblum_cnot_2018, xu_demonstration_2020}, as well as implementations of natively multi-mode error correcting codes, such as the pair-cat \cite{albert_pair-cat_2019}, Chuang-Leung-Yamamoto (CLY) \cite{Chuang_bosonic_1997}, or dual-rail codes \cite{chuang_simple_1995}. This may be done by complementing the dispersive control with tunable beamsplitter interactions \cite{gao_programmable_2018}, which swap states between oscillators, allowing the nonlinear ancilla to interact with each oscillator in turn. 

Recent progress in generating stronger tunable beamsplitter interactions between high-Q cavities without compromising their long coherence times or introducing unwanted nonlinearity \cite{chapman_high--off-ratio_2023, lu_high-fidelity_2023} provides access to a regime where the inter-oscillator coupling strength exceeds the typical coupling strength to the nonlinear ancilla. This presents the opportunity to treat the coupled oscillators collectively. Thus a single ancilla, statically coupled to only one of the modes, can be used to measure either joint or individual properties of the combined system. Such operations include the recently proposed joint-parity measurement or erasure check of a dual-rail cavity qubit \cite{tsunoda_error-detectable_2023, teoh_dual-rail_2023}.

In this article, we observe a new regime that emerges in the presence of an increasingly strong beamsplitter drive between two bosonic modes, which is an analog of dispersive number splitting \cite{schuster_resolving_2007}, but for the total excitation number, $N$, shared between two modes. In this regime the spectra of the combined multi-mode system are well predicted by treating the parametrically coupled oscillators as a single spin with $S=N/2$ that has a dispersive coupling to the ancilla. This provides an intuitive picture for arbitrary $N$. We find a range of operating points that are now accessible where we can measure the total photon number in the coupled oscillators, without measuring the photon number in either cavity.

An important example of an operation enabled in this regime is a mid-circuit erasure check \cite{levine_demonstrating_2024, koottandavida_erasure_2024}, an essential ingredient for a dual-rail qubit \cite{chuang_simple_1995} spanned by the single excitation manifold $\ket{0,1}$ and $\ket{1,0}$ of two superconducting cavity modes. Erasure qubits \cite{grassl_codes_1997, kubica_erasure_2023, koottandavida_erasure_2024, levine_demonstrating_2024, scholl_erasure_2023, wu_erasure_2022, kang_quantum_2023, ma_high-fidelity_2023, teoh_dual-rail_2023, chou_demonstrating_2023} rely on detecting dominant leakage errors and resetting these states back into the codespace. Since the time and location of these (erasure) errors are known, erasure qubits yield high thresholds when embedded in a higher-level error-correcting code \cite{barrett_fault_2010, delfosse_linear-time_2020, wu_erasure_2022}. The dual-rail cavity qubit is a prototypical example of this as its errors are dominated by detectable leakage to $\ket{0,0}$, with Pauli errors within the codespace much less likely \cite{teoh_dual-rail_2023}. 

We implement a minimally-invasive erasure check by only exciting the less-coherent ancilla transmon when the dual-rail qubit has already leaked from the codespace. This limits the probability of declaring an erasure due to errors in the measurement itself to $2.92\pm0.01\%$ per check, of which only $0.51\pm0.02\%$ are false positives due to transmon errors. With a missed-erasure probability of only $(9.0\pm0.5)\times 10^{-4}$, dual-rail logical states are also well-preserved, with a Pauli error rate of $0.31\pm0.01\%$, of which only $0.12\pm0.01\%$ is induced by transmon decoherence. As in the previous demonstration of mid-circuit erasure detection for a dual-rail cavity qubit by Koottandavida \textit{et al.} \cite{koottandavida_erasure_2024} this scheme engineers an ancilla spectrum that distinguishes between computational states and the joint vacuum state, $\ket{0,0}$. However, it does so by making use of the tunable beamsplitter already required for single- and two-qubit gates and minimizes the connectivity between circuit elements. Realizing such a  high-performance and minimally-invasive mid-circuit erasure check is a crucial capability for improving error correction \cite{wu_erasure_2022} via erasure conversion.

In the following sections, we begin by describing the emergence of parametrically-activated joint-photon number-splitting, before showing its application in a dual-rail erasure check. An extension of this technique to a two-qubit gate is described in Appendix~\ref{app:CPHASE}.

\section*{RESULTS}
\subsection*{Spectroscopic observation of joint-photon number-splitting}
\begin{figure*}[htb]
    \centering
    \includegraphics[width=1\linewidth]{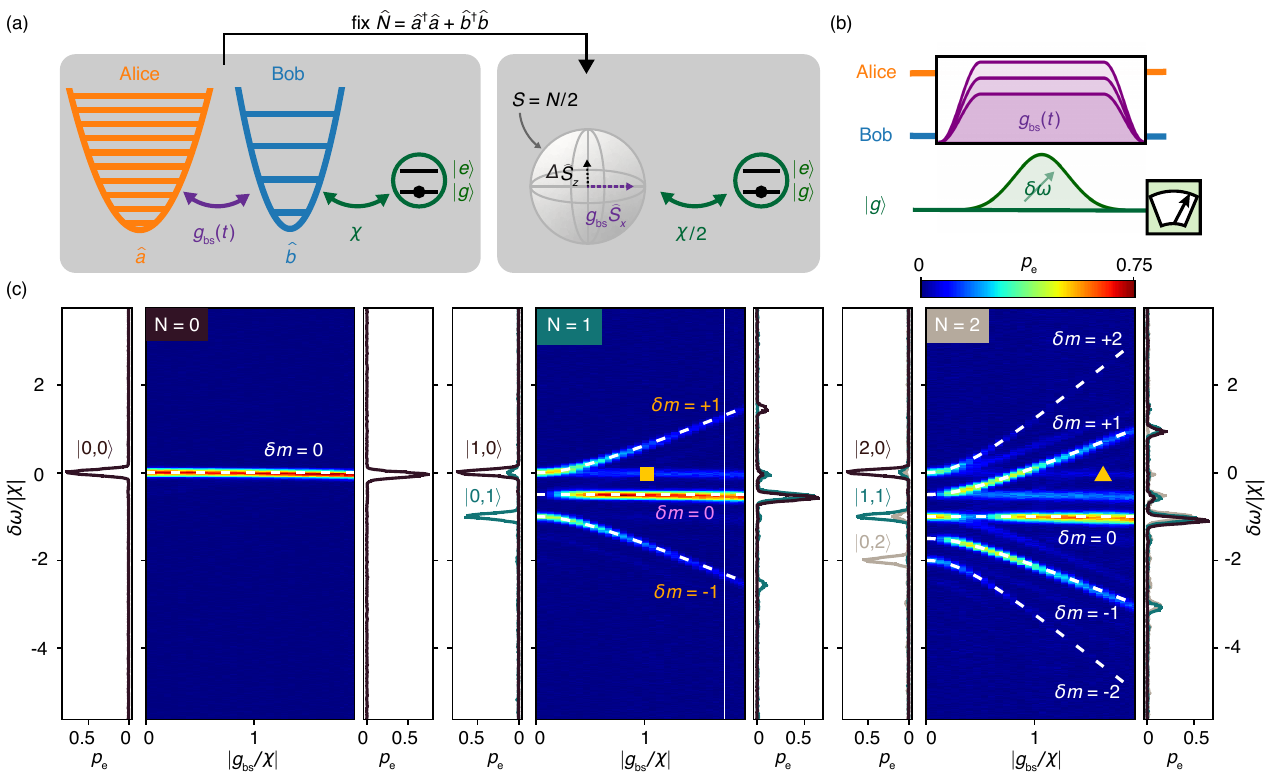}
    \caption{\textbf{Ancilla spectroscopy in the presence of a beamsplitter drive.} (a) System schematic showing two oscillators coupled by a tunable beamsplitter interaction and an ancilla qubit statically coupled to one oscillator via a fixed dispersive interaction. Two coupled oscillators with total photon number $N$ are equivalent to a spin with $S=N/2$. (b) Pulse sequence used for spectroscopy experiment. (c) Transmon spectra in the presence of increasing beamsplitter drive amplitude for input oscillator states with fixed total photon number $N$. Colorplots show initial states $\ket{0,0}$ ($N=0$), $\frac{\ket{0,1}+\ket{1,0}}{\sqrt{2}}$ ($N=1$) and $\frac{\ket{0,2} + \sqrt{2}\ket{1,1} + \ket{2,0}}{2}$ ($N=2$). Predicted transitions (dashed white lines) are labeled by the change in the magnetic quantum number $\delta m$ in the oscillator-spin model (see Fig.~\ref{fig:energy_levels_matrix_elements}). Left (right) panels show spectra for all initial two-oscillator Fock states in each $N$-photon manifold at the lowest (highest) value of $|g_{\mathrm{bs}}/\chi|$ shown in the colorplots. The yellow square (triangle) in the middle (right) panel indicates operating point for the erasure check (CPHASE gate described in Appendix~\ref{app:CPHASE}).}
    \label{fig:driven_spec}
\end{figure*}

Combining two common elements of oscillator control, a tunable beamsplitter interaction between two oscillators and a fixed dispersive interaction between a single oscillator and an ancillary qubit, enables measurements and control conditioned on the total photon number of the system, via an extension of the strong dispersive regime of circuit QED \cite{schuster_resolving_2007}.
The Hamiltonian of this system in the interaction picture is
\begin{equation}
    \frac{\hat{\mathcal{H}}}{\hbar} = \frac{g_{\text{bs}}(t)}{2}\left(e^{i\varphi}\hat{a}\hat{b}^{\dagger}+e^{-i\varphi}\hat{a}^{\dagger}\hat{b}\right) - \Delta \hat{b}^{\dagger}\hat{b} + \chi\hat{b}^{\dagger}\hat{b}\ket{e}\bra{e},
    \label{eqn:H}
\end{equation}
where the oscillator Alice and Bob modes are represented by lowering operators $\hat{a}$ and $\hat{b}$, and $\ket{e}$ is the excited state of the ancillary qubit (see Fig.~\ref{fig:driven_spec}a). The ancilla is coupled to only one of the modes with a dispersive interaction strength $\chi$ that is fixed, whereas the amplitude $g_{\mathrm{bs}}$, phase $\varphi$ and frequency detuning $\Delta$ of the beamsplitter drive are all controllable in time. The dispersive term (activated by intentionally exciting the ancilla out of its ground state) and the beamsplitter term are often used alternately: either narrow-bandwidth pulses on the ancilla enact photon-number-selective operations on a single mode or a beamsplitter routes states between modes \cite{gao_entanglement_2019}. However, when both terms are activated simultaneously and $g_{\mathrm{bs}}\sim \chi$, the number-split spectrum of the ancilla is modified to depend on the joint photon number in both oscillators.

The Hamiltonian in Eq.~\ref{eqn:H} is realized with a pair of superconducting microwave stub cavities as the oscillators \cite{reagor_quantum_2016} and a fixed-frequency transmon as the ancilla. The use of a SNAIL coupler \cite{frattini_3-wave_2017, frattini_three-wave_2021, chapman_high--off-ratio_2023, zhou2023realizing} situated between the cavities allows us to engineer a microwave-activated beamsplitter interaction with amplitude up to $|g_{\mathrm{bs}}|/2\pi = 2.05$~MHz, in excess of the dispersive coupling $|\chi|/2\pi = 1.07$~MHz, while preserving the coherence and linearity of the cavity modes.

The emergence of joint-photon number-splitting is revealed by probing the transmon spectrum in the presence of a variable-amplitude beamsplitter drive, as illustrated in Fig.~\ref{fig:driven_spec}b. Importantly, the beamsplitter drive is applied with a `symmetric' frequency detuning, $\Delta=\chi/2$ --- since Bob's frequency shifts by $\chi$ when the transmon is in $\ket{e}$, this condition ensures the beamsplitter drive is equally detuned from resonance when the transmon is in $\ket{g}$ or in $\ket{e}$. Because the Hamiltonian conserves the total oscillator photon number $N$ and the average cavity $T_1$ is much longer than the pulse duration $T_\mathrm{p}$, the dynamics may be considered separately for different values of $N$. Fig.~\ref{fig:driven_spec}c shows the resulting spectra for different initial states with $N=0, 1$ and $2$.

In the absence of a coupling between the cavities ($g_{\mathrm{bs}}=0$) the transmon spectrum displays the familiar photon number-splitting regime, with transitions separated in frequency by $\chi$ per photon in the Bob mode but independent of the photon number in the uncoupled Alice mode. This can be seen from the linecuts at $g_{\mathrm{bs}}=0$ (in the left panels of Fig.~\ref{fig:driven_spec}c) when initializing the cavities in different two-oscillator Fock states. However, in the presence of a strong beamsplitter drive $\left(|g_{\mathrm{bs}}|\sim |\chi|\right)$ we enter the joint-photon number-splitting regime. Here, the transmon spectrum exhibits $2N+1$ prominent transition lines, with the central transition at a frequency detuning $\delta\omega = N\chi/2$ becoming the dominant one at large values of $g_{\mathrm{bs}}$.

We can compare these features to analytical predictions obtained using Schwinger's angular momentum operator formalism \cite{schwinger_angular_1952, tsunoda_error-detectable_2023}, which considers the two cavities containing $N$ total photons as a single spin with $S=N/2$ (see Fig.~\ref{fig:driven_spec}a). This formalism provides a way to map oscillator operators onto spin operators:
\begin{align}
\begin{aligned}
    \hat{S}_x \equiv \frac{\hat{a}^{\dagger}\hat{b} + \hat{b}^{\dagger}\hat{a}}{2}, \\
    \hat{S}_z \equiv \frac{\hat{a}^{\dagger}\hat{a} - \hat{b}^{\dagger}\hat{b}}{2},
\end{aligned}
&&
\begin{aligned}
    \hat{S}_y \equiv \frac{\hat{a}^{\dagger}\hat{b} - \hat{b}^{\dagger}\hat{a}}{2i}, \\ \frac{N}{2}\hat{\mathbb{I}} \equiv \frac{\hat{a}^{\dagger}\hat{a} + \hat{b}^{\dagger}\hat{b}}{2}.
\end{aligned}
\label{eqn:schwinger}
\end{align}
For a symmetrically-detuned beamsplitter drive ($\Delta=\chi/2$) this gives a spin Hamiltonian for this system:
\begin{equation}
    \frac{\hat{\mathcal{H}}}{\hbar} =\ g_{\text{bs}}\hat{S}_x+\frac{\chi}{2}\hat{S}_z\hat{\sigma}_z -\frac{N\chi}{4}\hat{\sigma}_z,
\end{equation}
where $\hat{\sigma}_z=\ket{g}\bra{g}-\ket{e}\bra{e}$ is the Pauli-Z operator for the ancilla and the beamsplitter drive phase is chosen to be $\varphi = 0$. When the amplitude of the spectroscopy drive on the qubit is small relative to $\text{max}(|g_{\mathrm{bs}}|, |\chi|)$, its effect in the Hamiltonian may be treated as a perturbation. $\hat{\mathcal{H}}$ is then block-diagonal with respect to the ancilla subspace, allowing us to separately consider Hamiltonians conditioned on the ancilla state, $\hat{\mathcal{H}}_g \equiv \bra{g}\hat{\mathcal{H}}\ket{g}$ and $\hat{\mathcal{H}}_e \equiv \bra{e}\hat{\mathcal{H}}\ket{e}$:
\begin{equation}
    \hat{\mathcal{H}} = \hat{\mathcal{H}}_g\ket{g}\bra{g} + \hat{\mathcal{H}}_e\ket{e}\bra{e}.
    \label{eqn:diagonal}
\end{equation}
Diagonalizing $\hat{\mathcal{H}}_g$ and $\hat{\mathcal{H}}_e$ gives spin eigenstates (representing the state of the coupled oscillators) before and after the ancilla transition from $\ket{g}$ to $\ket{e}$, with each Hamiltonian defining a spin quantization axis along which these eigenstates are aligned.

\begin{figure}[t!]
    \centering
    \includegraphics[width=1\linewidth]{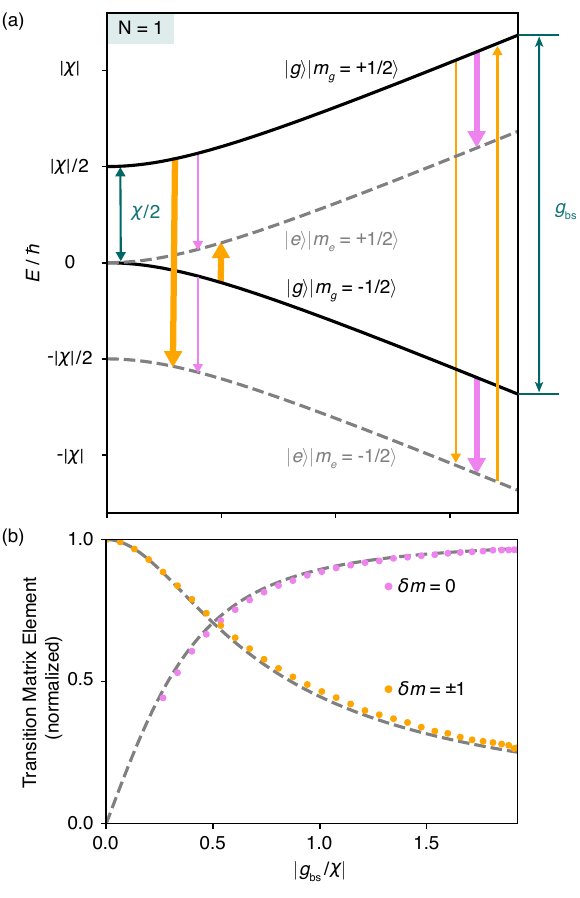}
    \caption{\textbf{Ancilla transition matrix elements in the presence of beamsplitter.} (a) Energy level diagram for the ancilla-oscillator system with $N=1$ photon shared between the two oscillators. States are labeled by their transmon state ($\ket{g}$ or $\ket{e}$) and oscillator-spin state (magnetic quantum number $m_{g/e}=\pm1/2$).  Arrow thicknesses illustrate the strength of transition matrix elements at low and high beamsplitter amplitudes. (b) Measured transition matrix elements along central ($\delta m=0$, pink) and lower ($\delta m=-1$, orange) transitions of $N=1$ spectrum, obtained from ancilla Rabi oscillation rates. Dashed lines show predictions for normalized transition matrix elements obtained from the spin model.}
    \label{fig:energy_levels_matrix_elements}
\end{figure}

In this spin picture, the application of the beamsplitter drive is analogous to a transverse magnetic field. As the beamsplitter strength increases, the energy eigenstates are no longer oscillator Fock states ($\hat{S}_z$ eigenstates) but more closely resemble symmetric and antisymmetric superpositions of these Fock states ($\hat{S}_x$ eigenstates). Meanwhile the eigenenergies, shown for total photon number $N=1$ in Fig.~\ref{fig:energy_levels_matrix_elements}a, display a zero-field splitting by $\chi/2$ and a high-field Zeeman-like splitting that approaches $g_{\mathrm{bs}}$ as the transverse field is increased (i.e. by increasing the beamsplitter strength). The eigenstates are labelled by their projection along the quantization axis:
\begin{align}
    \hat{\mathcal{H}}_{g} \ket{m_{g}} &= \left(-\frac{N\chi}{4} + m_{g}\hbar\, \Omega\right)  \ket{m_{g}}, \\
    \hat{\mathcal{H}}_{e} \ket{m_{e}} &= \left(+\frac{N\chi}{4} + m_{e}\hbar\, \Omega\right)  \ket{m_{e}},
\end{align}
where the magnetic quantum numbers can take the values,
\begin{equation}
    m_{g/e}=-S, \dots, S=-\frac{N}{2},\dots,\frac{N}{2}, \nonumber
\end{equation}
and $\Omega$ can be interpreted as a Larmor frequency:
\begin{equation}
\Omega=\sqrt{g_{\mathrm{bs}}^2+\left(\frac{\chi}{2}\right)^2}.
\end{equation}

The energy difference between each possible pair of $\hat{\mathcal{H}}_g$ and $\hat{\mathcal{H}}_e$ eigenstates allows us to accurately predict the $2N+1$ unique transition frequencies observed in the spectra:
\begin{equation}
    \omega_{\delta m} = \frac{N\chi}{2}+\delta m\,\Omega, \quad\delta m\equiv m_e-m_g=-N,\ldots,N.
    \label{eqn:freqs}
\end{equation}
These frequencies are shown as white dashed lines in Fig.~\ref{fig:driven_spec}c, and show good agreement with the observed spectra \footnote{We note that the spectrum for $N=1$ is reminiscent of the Mollow triplet \cite{mollow_power_1969, baur_measurement_2009} seen when probing the frequency of a strongly-driven two-level-system. Here we are instead probing the frequency of a two-level-system that is dispersively coupled to the strongly driven mode.}. Faint features corresponding to transitions in the $N-1$ photon manifold can also be seen due to photon loss during the spectroscopy pulse. The symmetric beamsplitter detuning condition $\Delta=\chi/2$ provides a unique operating point where each transition frequency has a degeneracy $(N+1)-|\delta m|$.  (In Appendix \ref{app:degeneracy_breaking}, we show how shifting away from this point breaks the degeneracies between these transitions, yielding $(N+1)^2$ unique frequencies.)

Furthermore, the strength of each transition is proportional to the overlap between the initial ($\hat{\mathcal{H}}_g$) and final ($\hat{\mathcal{H}}_e$) eigenstates. As $g_{\mathrm{bs}}$ is increased, both sets of eigenstates start to align along the $x$-axis, with increasing overlap between states where $m_g = m_e$. As a result, off-central transitions with $\delta m \neq 0$ become suppressed in favor of transitions with $\delta m=0$, as can be seen from their changing brightness in Fig.~\ref{fig:driven_spec}c. We verify this quantitatively for transitions in the $N=1$ manifold by measuring the rate of ancilla Rabi oscillations across a range of $|g_{\mathrm{bs}}/\chi|$ values (described in Appendix ~\ref{app:matrix_elements}). The measured rates for the $\delta m=0$ and $\delta m=\pm1$ transitions (proportional to the transition matrix element) are shown in Fig.~\ref{fig:energy_levels_matrix_elements}b and agree well with the model predictions.

The predictions from this spin model explain the emergence of a joint-photon number-splitting regime when $g_{\mathrm{bs}}> \chi$, where the qubit spectrum depends on the total number of photons in both oscillators. This is most clear in linecuts of the spectra at the largest values of $g_{\mathrm{bs}}/\chi$ in Fig.~\ref{fig:driven_spec}c, where the central $\delta m=0$ transition line dominates. Independent of which state we initialize in the oscillators, we see a single dominant transition frequency at $\omega_0 = N\chi/2$ for each value of $N$. This can be interpreted as the indistinguishability of photons in Alice and Bob from the perspective of the qubit when the beamsplitter interaction is stronger than the dispersive shift. The extent to which the qubit can still distinguish the two oscillators can be seen in the much smaller satellite $\delta m=\pm 1$ peaks, at a detuning of $\pm g_{\mathrm{bs}}$, which would be further suppressed at higher beamsplitter amplitudes.

In the following section we show how understanding these joint-photon number-split spectra provides a way to extend established single-oscillator control techniques to operations on multiple oscillators.

\subsection*{Mid-circuit erasure check of a dual-rail qubit}

\begin{figure*}[ht]
    \centering
    \includegraphics[width=1\linewidth]{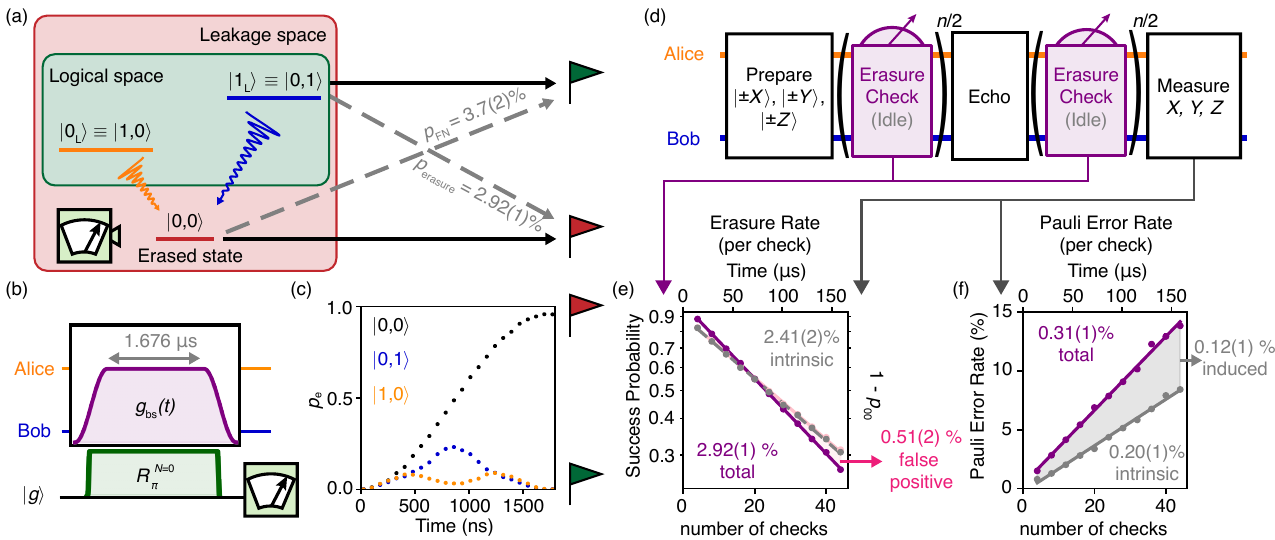}
    \caption{\textbf{Characterizing a dual-rail mid-circuit erasure check.} (a) Logical code space of a dual-rail encoded qubit indicating decay to the leakage state $\ket{0,0}$, which we seek to detect and convert to erasure errors. Black arrows show the ideal mapping processes shown while gray arrows indicate misassignment processes that lead to false negative (FN) errors and erasure errors respectively. We measure a false negative error rate of $p_{\mathrm{FN}} = 3.7(2)\%$ per erasure check and a total erasure error rate of $p_{\mathrm{erasure}} = 2.92(1)\%$ per erasure check. Green (red) flag indicates erasure check reporting ``no erasure'' (``erasure detected''). (b) Pulse sequence used to implement the erasure check, consisting of simultaneous flat-top pulses on both the transmon and the beamsplitter drive, followed by transmon readout. (c) Measured transmon trajectories during the erasure check for logical and leakage states. (d) Pulse sequence used to measure state fidelity after $n$ repeated erasure checks (or $n$ periods of idling for the duration of an erasure check) on the encoded dual-rail qubit. A single echo pulse (a cavity SWAP operation) is performed halfway through the sequence in order to suppress the effect of no-jump backaction resulting from different cavity decay rates. (e) Success probability for passing $n$ consecutive checks (purple), and probability of remaining in logical subspace ($1-p_{00}$) after $n$ checks, unconditioned on erasure check results, with (gray) and without (pink) the beamsplitter pump applied. (f) Total Pauli error rate when performing erasure checks (purple) or idling (gray). Lines show linear fits.}
    \label{fig:Erasure_Check}
\end{figure*}
We use the $|g_{\mathrm{bs}}| \gtrsim |\chi|$ regime to construct a mid-circuit erasure check for a dual-rail qubit encoded in two superconducting cavities. The basis states for this encoding are  $\ket{0_\text{L}} \equiv \ket{1,0}$ and $\ket{1_\text{L}} \equiv \ket{0,1}$ and their dominant error channel is photon loss to the common $\ket{0,0}$ leakage state, which we hope to detect (see Fig.~\ref{fig:Erasure_Check}a). Detecting these leakage errors (in order to convert them to erasures) while preserving logical information in states that have not suffered such an error is an essential task in this architecture.

A key requirement of the erasure check is that it does not introduce additional errors. Firstly, since the check may be performed multiple times per round of syndrome measurements, erasure errors during the check must be minimized for the code to operate below the error-correction threshold of the higher-level code, such as the surface code \cite{kitaev_fault-tolerant_2003, dennis_topological_2002, fowler_high-threshold_2009}. Secondly, Pauli errors must remain much less likely than erasure errors to preserve the bias that enables a high threshold \cite{wu_erasure_2022}. Both of these criteria require a way of limiting errors induced by the less-coherent ancilla transmon.

A powerful and established technique for measuring bosonic modes while preventing the ancilla from polluting the logical state is to use a three-level ancilla and to apply a microwave drive to ensure that the dispersive shift is unchanged when the ancilla is in $\ket{e}$ or in $\ket{f}$ (often known as $\chi$-matching) \cite{Rosenblum_fault-tolerant_2018}. This erasure check, however, bypasses the need for $\chi$-matching drives since it only needs to catch events when the logical information has already been lost.

The modification of the transmon spectrum in the presence of a beamsplitter drive allows us to perform an erasure check with low susceptibility to transmon errors. In the number-splitting regime for a single oscillator, a narrow-bandwidth pulse on the ancilla enables a $\pi$-pulse conditioned on zero photons in the oscillator. Likewise, the joint-photon number-splitting regime allows us to excite the transmon if and only if there are zero total photons in two oscillators (i.e. when the dual-rail qubit has leaked to $\ket{0,0}$).

For transmon pulses selective on $N=0$, we do not require $g_{\mathrm{bs}}\gg\chi$ but only that the $N=0$ transition is sufficiently detuned from all other transitions we would like to avoid. In the context of the dual-rail encoding, these are the transitions in the $N=1$ manifold (corresponding to the logical subspace), for which the detuning saturates at $\chi/2$ for $|g_{\mathrm{bs}}/\chi|\geq \sqrt{3}/2$ (see Fig.~\ref{fig:driven_spec}c). This establishes a wide range of values for $|g_{\mathrm{bs}}/\chi|$ that can enact an erasure check and allows for a more flexible implementation than the erasure check proposed in \cite{tsunoda_error-detectable_2023, teoh_dual-rail_2023}, based on measuring the joint photon number parity, where $|g_{\mathrm{bs}}/\chi|$ is fixed (see Appendix~\ref{app:joint_parity} for a detailed comparison of the two approaches).

Furthermore, a variety of pulse shapes on the transmon drive may be used to perform the erasure check while trading off susceptibility to different error sources. For example, a long, highly frequency-selective Gaussian pulse will limit transmon excitation when in the $N=1$ manifold, at the expense of more idling errors during the check. To minimize the combined rate of transmon errors and idling errors, we use a shorter square pulse with a duration $T_{\text{p}} = 1.820~\muup$s while applying a beamsplitter drive with amplitude $|g_{\mathrm{bs}}/\chi|= 1.04$ (as indicated by the square symbol in Fig.~\ref{fig:driven_spec}c). These values of $g_{\text{bs}}$ and $T_{\text{p}}$ are calibrated to ensure that when starting in the dual-rail logical subspace, both the transmon and oscillators return to their initial states at the end of the sequence despite the finite selectivity of the transmon pulse (see Fig.~\ref{fig:Erasure_Check}b,c; for detailed calibration procedures, see Appendix~\ref{app:tuneup}). While this scheme places no special requirement on the adiabaticity of the beamsplitter drive ramp, we use a 120~ns cosine-shaped ramp to stay well within the bandwidth limit imposed by the on-chip Purcell filter through which the beamsplitter drive is applied. We align the center of this ramp in time with the center of the 24~ns cosine ramp on the transmon pulse, which we find from simulation ensures optimal performance.

The trajectories of the transmon state for different input oscillator states in Fig.~\ref{fig:Erasure_Check}c showcase the operation of the erasure check: while the transmon ends in $\ket{e}$ when the oscillators are in $\ket{0,0}$ (thus flagging an erasure), it returns to $\ket{g}$ for input states $\ket{0,1}$ and $\ket{1,0}$. The relatively small area under the $\ket{0,1}$ and $\ket{1,0}$ curves indicates that the transmon is less likely to be excited in these cases and so transmon decay and dephasing errors are less likely to induce false positives or logical Pauli errors. Meanwhile the relatively small difference between the $\ket{0,1}$ and $\ket{1,0}$ curves indicates the limited extent to which transmon dephasing errors allow the environment to distinguish between different logical states, thereby inducing Pauli errors when they occur \cite{ma_path-independent_2020}.

We test the fraction of leakage errors caught by the check by preparing $\ket{0,0}$ and performing a single mid-circuit erasure check followed by destructive photon-number-selective measurements of each cavity \cite{levine_demonstrating_2024}. When post-selecting on the final state remaining in $\ket{0,0}$, we find a false negative rate $p_{\mathrm{FN}} = 3.7 \pm 0.2\%$, consistent with results of master equation simulations using the physical transmon error rates in our system. While the false negative rate \textit{is} sensitive to transmon errors, in the operation of an error-correcting code the actual probability of missing an erasure $(p_{\mathrm{miss}} = p_{\mathrm{erasure}}\times p_{\mathrm{FN}}\sim 10^{-3}$) is also multiplied by the  small probability that an erasure has been suffered since the previous check. We find that this rate of false negatives is still admissible for high fault-tolerant thresholds in the surface code. In fact, when performing erasure checks after every two qubit gate, the surface code threshold (per step consisting of a gate plus an erasure check) with such a level of false negatives is $p_{\text{th}} = 3.71 \pm 0.02 \%$, which is well in excess of the $1\%$ Pauli noise threshold. As a point of comparison, if the false negative rate were 0, our threshold would be $p_{\text{th}} = 3.79 \pm 0.02\%$, signifying that the threshold is minimally affected by the rate of false negatives. In both cases, we have assumed that $90\%$ of errors are erasures. For more details on these simulations, refer to Appendix~\ref{app:thresholdsimulations}. As such, false negatives, which have a very small impact on $p_{\text{th}}$, are less costly than false positives, which contribute to the overall physical error rate that should be kept small with respect to $p_{\text{th}}$. This informs the design of the erasure check where the transmon is excited only if $N=0$.

A single experiment is used to evaluate the performance of the mid-circuit erasure check when acting on the code space, in terms of both the erasure error rate and the Pauli error rate induced by the check. This tomography sequence, shown in Fig.~\ref{fig:Erasure_Check}d, consists of preparing the six dual-rail cardinal states $\ket{\pm X},\ket{\pm Y}$, and $\ket{\pm Z}$, repeating the erasure check $n$ times and then measuring the logical operators $\hat{X}_\text{L}, \hat{Y}_\text{L}$ and $\hat{Z}_\text{L}$ using photon-number-selective readout of each cavity (see Appendix~\ref{app:EOL}). To separate the contribution from idling errors, we perform the same sequence but replacing each erasure check with a delay of the same duration. Beamsplitter pulses are used to enact single-qubit gates on the dual-rail qubit, allowing us to prepare states on the equator of the Bloch sphere and perform logical measurements. An echo pulse is added to remove the effect of no-jump back-action at long times. When the decay rate in each cavity is different (as is especially the case here, with $T_{1,a} = 347~\muup$s and $T_{1,b} = 109~\muup$s), post-selecting on no photon loss leads to a deterministic polarization towards one pole of the Bloch sphere and results in an approximately-Gaussian envelope on both the idling and erasure check data on a timescale set by the difference in the $T_1$ times \cite{teoh_dual-rail_2023}. Introducing the echo therefore allows us to better resolve errors induced by the check itself.

The erasure rate is extracted by looking at the success probability (i.e. the likelihood of passing $n$ successive checks) as a function of $n$, averaged over all input states (see Fig.~\ref{fig:Erasure_Check}e). The slope of the exponential decay shows a total erasure rate per check $p_{\text{erasure}} = 2.92 \pm 0.01\%$. To determine what fraction of these flagged events are `intrinsic' erasures due to photon loss to $\ket{0,0}$, as opposed to false positive events predominantly due to transmon errors, we may instead ignore the results of the mid-circuit checks and only ask how often the end-of-line measurement yields $\ket{0,0}$. This gives a photon loss rate of $2.41 \pm 0.02\%$ per check, consistent with the value obtained when idling for the same duration, $2.43 \pm 0.02\%$, indicating that the erasure check does not induce additional photon loss in the cavities. This value is the `intrinsic' erasure rate, from which we infer that the remaining $0.51 \pm 0.02\%$ of erasures are false positive errors. That the majority of detected erasures are due to photon loss in the high-Q cavities, rather than transmon errors, shows the ability of the scheme to tolerate decoherence in the transmon ancilla. Furthermore, the flexibility to choose different transmon pulse shapes allows us to trade off between `intrinsic erasures' (suppressed with a shorter pulse) and false positives (suppressed with a longer, more selective pulse), which can therefore be tailored to the relative decoherence rates in the system. This flexibility, not present in the joint-parity-based approach \cite{tsunoda_error-detectable_2023}, also allows us to trade off false positives ($p_{\text{FP}}=0.51\%$) for less-costly false negatives ($p_{\text{FN}}=3.7\%$).

To verify that this mid-circuit erasure check preserves the logical qubit state, we probe the fidelity of all 6 dual-rail cardinal states conditioned on passing $n$ successive erasure checks. Fitting the slope of the post-selected Pauli error probabilities (obtained from the average state fidelity; see Appendix~\ref{app:QPT}) as a function of measurement rounds $n$ allows us to precisely resolve the error rate for a single round (Fig.~\ref{fig:Erasure_Check}f). We find that the logical information is well-preserved during the course of the measurement, with an overall Pauli error per check $p_{\text{Pauli}} = 0.31 \pm 0.01\%$, compared to $0.20 \pm 0.01\%$ when idling. This indicates that background cavity errors dominate, with transmon-induced Pauli errors contributing at most the remaining $0.12 \pm 0.01\%$,  highlighting the robustness of the scheme against transmon decoherence. We note that while the echo pulse used to mitigate against no-jump backaction will also reduce the effect of low-frequency dephasing noise on the intrinsic error rate, errors induced by transmon decoherence should not be affected.

Taken together, these results demonstrate the efficacy of a flexible hardware-efficient mid-circuit erasure check, making use of only the beamsplitter interaction (used for gates) and dispersive transmon coupling to a single mode (used for state preparation). This check preserves a large ratio of the erasure error rate to the Pauli error rate, with both quantities remaining dominated by `intrinsic' errors associated with the hardware. With improvement of the mode coherences closer to state-of-the-art values \cite{place_new_2021, ganjam_surpassing_2024,oriani_niobium_2024}, we expect this scheme to yield significantly below-threshold performance. Indeed, with the same $\chi$, $g_{\mathrm{bs}}$ and readout duration $\tau_{\text{RO}}$ but with a transmon $T_1$ and $T_\phi$ of $200~\muup \text{s}$ and an average cavity $T_1$ of $1000~\muup \text{s}$, we predict from master equation simulations a total erasure rate $p_{\mathrm{erasure}} = 0.49\%$ and a transmon-induced Pauli error rate of $0.035\%$, at which point shot-noise dephasing from photons in the readout resonator starts playing a larger role (see Appendix~\ref{app:expected_performance}). Increasing $\chi$ and reducing $\tau_{\text{RO}}$ to reduce the check duration provides another way to improve performance. Separately, increasing $g_{\mathrm{bs}}$ would help reduce Pauli errors from transmon dephasing by making the dual-rail logical states more indistinguishable during the check \cite{ma_path-independent_2020} (see Appendix~\ref{app:error_scaling}).

\section*{DISCUSSION}
These results show that a strong beamsplitter drive, when paired with dispersive coupling to a single mode, offers a versatile means of measuring and controlling multiple oscillators. The tunable beamsplitter allows for switching between single-cavity operations (accessible when $g_{\mathrm{bs}}=0$) and their equivalent multi-cavity operations on the total photon number (accessible when $g_{\mathrm{bs}}\gtrsim\chi$). As an important example of this, we have demonstrated how photon-number selective measurements can be extended to joint-photon-number selective measurements to enable a mid-circuit erasure check for dual-rail qubits. Further examples of this principle include the extension of the ``selective number arbitrary phase'' (SNAP) gate to a joint-SNAP-like gate selective on the photon number in two cavities, enabling a tunable CPHASE($\theta$) gate for two dual-rail qubits (see Appendix~\ref{app:CPHASE}).

The mid-circuit erasure check for dual-rail qubits enabled by joint-photon-number selective measurements represents a new, hardware-efficient way of performing the essential ingredient for an error-correcting surface code with dual-rail cavity qubits. The design of this check, which asks whether the joint photon number is zero (as opposed to a joint-parity check which asks whether it is even), leverages two important features of erasure errors: that they are rare, and that the state need not be preserved once an erasure is detected. Therefore by only minimally exciting the ancilla when in the dual-rail code space, it ensures that the contribution of transmon-induced erasures and Pauli errors is subdominant, at the cost of more false negatives, to which the code is more tolerant. One caveat is that this check does not catch rare but damaging heating events, although this could be mitigated with the addition of a selective pulse acting on the two-photon manifold.

Improvement of the mode coherences to state-of-the-art values should enable performance substantially below the erasure threshold for the surface code. With high-fidelity state preparation, logical erasure-detected measurements, single-qubit gates and hardware-efficient mid-circuit erasure checks demonstrated, key next steps will include a fully error-detectable two-qubit gate, as well as fast qubit reset to turn leakage detection into erasure conversion.

\section*{METHODS}

\subsection*{Device}
The device used for this experiment was previously used in Chapman et al. \cite{chapman_high--off-ratio_2023} and includes two superconducting $\lambda/4$ post cavities machined from 99.999\%-purity Aluminum as the two oscillators. Into 
each of these cavities we insert an EFG sapphire chip supporting a transmon qubit, a readout resonator and a Purcell filter. These are used to prepare and readout states in the two oscillators. The transmon coupled to Bob's cavity is used to determine the spectrum in Fig.~\ref{fig:driven_spec} and to perform the erasure check in Fig.~\ref{fig:Erasure_Check}. Detailed characterization of system parameters are shown in Table~\ref{tab:sys_params} and discussed in Appendix~\ref{app:system_params}. 

To provide the beamsplitter interaction between these cavities, we add a capacitively-shunted superconducting nonlinear asymmetric inductive element (SNAIL) \cite{frattini_three-wave_2021}, a superconducting loop with three nominally-identical Josephson junctions (each with Josephson energy $E_{\text{J}}/h=90.0\pm0.3$~GHz) in series on one arm of the loop and a single junction (with Josephson energy $\alpha=0.147\pm0.001$ times smaller) on the other. The capacitive shunt is implemented by adding leads to the two ends of the SNAIL connecting to large capacitor pads. This gives the SNAIL a charging energy $E_{\text{C}}/h=177\pm2$~MHz, while the leads also contribute a series linear inductance with energy $E_{\text{L}}/h=64\pm2$~GHz. This circuit is patterned onto a sapphire chip which is inserted into a tunnel that intersects both cavities, thereby generating a linear coupling between the SNAIL mode and each of the oscillator modes. Passing a DC magnetic flux through the SNAIL loop (delivered via a superconducting flux transformer) modifies the potential of the SNAIL, allowing us to generate a third-order nonlinearity. By applying a single microwave drive that couples to the charge operator of the SNAIL mode, at a frequency equal to the difference between the two cavity frequencies, we can generate the beamsplitter interaction via a three-wave-mixing process. The frequency detuning from this resonance condition is the detuning $\Delta$ indicated in Eq.~\ref{eqn:H}, and the amplitude and phase of the drive determine $g_{\mathrm{bs}}$ and $\varphi$.

Any magnetic flux bias $\Phi_{\text{ext}}$ besides a half integer multiple of $\Phi_0$ gives the SNAIL a third-order nonlinearity and at low drive powers, the magnitude of this nonlinearity determines the ratio between the applied drive amplitude and $g_{\mathrm{bs}}$. However, as was seen previously \cite{chapman_high--off-ratio_2023}, at larger drive powers, this linear relationship breaks down and there exists a maximum value of $g_{\mathrm{bs}}$ at each $\Phi_{\text{ext}}$. The flux bias used in this experiment ($\Phi_{\text{ext}}=0.334\Phi_0$) was chosen to give the highest maximum value of $g_{\mathrm{bs}}$ while avoiding any unwanted resonances at any value of $g_{\mathrm{bs}}$ as it is increased to this maximum value. This therefore provides access to the widest range of values for $|g_{\mathrm{bs}}/\chi|$. Appendix \ref{app:calibration} provides details on how we measure and calibrate the value of $g_{\mathrm{bs}}$.

The aluminum package is mounted to the mixing chamber plate of a Bluefors XLD400sl dilution refrigerator, at a temperature of 8~mK.

\subsection*{Oscillator-spin model}
As in Tsunoda et al. \cite{tsunoda_error-detectable_2023}, we can express the system Hamiltonian in Eq.~\ref{eqn:H} in terms of the Schwinger angular momentum operators \cite{schwinger_angular_1952} (defined in Eq.~\ref{eqn:schwinger}):
\begin{align}
    \frac{\hat{\mathcal{H}}}{\hbar} =\ &g_{\mathrm{bs}}\cos\varphi\hat{S}_x - g_{\mathrm{bs}}\sin\varphi\hat{S}_y \ - \nonumber \\
    &\left(\Delta-\chi\ket{e}\bra{e}\right)\left(\frac{\hat{N}}{2}-\hat{S}_z\right).
\end{align}
As indicated in Eq.~\ref{eqn:diagonal}, $\hat{\mathcal{H}}$ is block-diagonal in the ancilla subspace, allowing us to separately consider the ancilla-state-dependent Hamiltonians $\hat{\mathcal{H}}_g$ and $\hat{\mathcal{H}}_e$.

This Hamiltonian also conserves total photon number $\hat{N}$ and so if we restrict ourselves to the $N$-photon manifold, we may make the replacement $\hat{N} \rightarrow N$. In the $N$-photon manifold, the Schwinger angular momentum operators defined above are the standard spin-$N/2$ operators, with eigenvalues running from $-N/2$ to $N/2$. We may combine these spin operators into a single spin operator $\hat{S}_{\hat{\Omega}}$ pointing along a quantization axis $\hat{\Omega}$ such that the Hamiltonian is diagonal. Writing the Hamiltonian separately when the qubit is in $\ket{g}$ and in $\ket{e}$ gives:
\begin{align}
    \frac{\hat{\mathcal{H}}_g}{\hbar} &= \frac{-N\Delta}{2} + |\vec{\Omega}_g|\hat{S}_{\hat{\Omega}_g}, \\
    \frac{\hat{\mathcal{H}}_e}{\hbar} &= \frac{N\left(\chi-\Delta\right)}{2} + |\vec{\Omega}_e|\hat{S}_{\hat{\Omega}_e},
\end{align}
where 
\begin{align}
    \vec{\Omega}_g &= 
    \begin{pmatrix}
        g_{\mathrm{bs}}\cos\varphi \\
        -g_{\mathrm{bs}}\sin\varphi \\
        \Delta
    \end{pmatrix}, \\
    \vec{\Omega}_e &= 
    \begin{pmatrix}
        g_{\mathrm{bs}}\cos\varphi \\
        -g_{\mathrm{bs}}\sin\varphi \\
        \Delta-\chi
    \end{pmatrix}.
\end{align}

\subsubsection*{Transition frequencies}
Since the Hamiltonian is diagonal in each case, we may simply read off the eigenenergies as
\begin{align}
    \left\{\frac{E_g}{\hbar}\right\} &= \left\{-\frac{N\Delta}{2} + m_g\sqrt{g_{\mathrm{bs}}^2+\Delta^2}\right\}, \\
    \left\{\frac{E_e}{\hbar}\right\} &= \left\{\frac{N\left(\chi-\Delta\right)}{2}+m_e\sqrt{g_{\mathrm{bs}}^2+\left(\Delta-\chi\right)^2}\right\},
\end{align}
where $m_g, m_e = -N/2, ..., N/2$. If we take all the differences between these energies, we obtain $(N+1)^2$ transition frequencies,
\begin{align}
    \left\{\omega\right\} = \bigg\{\frac{N\chi}{2} + &m_e\sqrt{g_{\mathrm{bs}}^2+\left(\Delta-\chi\right)^2} \nonumber \\
    - &m_g\sqrt{g_{\mathrm{bs}}^2+\Delta^2}\bigg\}.
    \label{eqn:nondegen_freqs}
\end{align}

The case where $\Delta=\chi/2$, the condition required to observe joint-photon number-splitting, is special as many of the transition frequencies become degenerate, leaving $2N+1$ unique transitions:
\begin{equation}
    \left\{\omega\right\}_{\Delta=\chi/2} = \left\{\frac{N\chi}{2} + \delta m\sqrt{g_{\mathrm{bs}}^2+\left(\frac{\chi}{2}\right)^2}\right\}
\end{equation}
where $\delta m \equiv m_e - m_g = -N, ..., N$.

\subsubsection*{Transition matrix elements}
From the eigenstates of $\mathcal{H}_g$ and $\mathcal{H}_e$, which we denote as $\ket{m_g}$ and $\ket{m_e}$, we may also find the matrix elements for each transition:
\begin{align}
    |M|_{m_g\rightarrow m_e} &= \left|\bra{m_g}\bra{g}\left(\frac{\epsilon}{2} \ket{g}\bra{e} + \frac{\epsilon^*}{2}\ket{e}\bra{g}\right)\ket{e}\ket{m_e}\right| \nonumber\\
    &= \frac{|\epsilon|}{2}\left|\braket{m_g}{m_e}\right|
\end{align}
The overlap between the initial and final oscillator states therefore tells us the strength of each transition. The overlap between a spin-$N/2$ eigenstate with magnetic quantum number $m_g$ along one axis and another eigenstate with magnetic quantum number $m_e$ along another axis at an angle $\delta\theta$ from the first is given by the Wigner (small) d-matrix \cite{wigner_gruppentheorie_1931}:
\begin{equation}
    \left|\braket{m_g}{m_e}\right| = \left|d_{m_g,m_e}^{N/2}(\delta\theta)\right|.
\end{equation}
In our case, the angle between the spin axes $\hat{\Omega}_g$ and $\hat{\Omega}_e$ is
\begin{equation}
    \delta\theta = \arctan\left(\frac{\Delta}{g_{\mathrm{bs}}}\right) - \arctan\left(\frac{\Delta-\chi}{g_{\mathrm{bs}}}\right).
\end{equation}
\smallskip

For the specific case shown in Fig.~\ref{fig:energy_levels_matrix_elements}b, where $N=1$ and $\Delta=\chi/2$, the angle difference becomes $\delta\theta=2\arctan\left(\chi/2g_{\mathrm{bs}}\right)$. The matrix element for the central transition is therefore
\begin{align}
    \left|M_{\delta m = 0}\right| &\propto \left|d_{\frac{1}{2}, \frac{1}{2}}^{\frac{1}{2}}(\delta\theta)\right| =  \left|d_{-\frac{1}{2}, -\frac{1}{2}}^{\frac{1}{2}}(\delta\theta)\right| \\
    &=\cos\left(\frac{\delta\theta}{2}\right) \\
    &= \cos\left(\arctan\left(\frac{\chi}{2g_{\mathrm{bs}}}\right)\right) \\
    &= \frac{g_{\mathrm{bs}}}{\sqrt{g_{\mathrm{bs}}^2+\left(\frac{\chi}{2}\right)^2}}.
\end{align}
Likewise for the off-central transitions,
\begin{align}
    \left|M_{\delta m = \pm1}\right| &\propto \left|d_{\frac{1}{2}, -\frac{1}{2}}^{\frac{1}{2}}(\delta\theta)\right| =  \left|d_{-\frac{1}{2}, \frac{1}{2}}^{\frac{1}{2}}(\delta\theta)\right| \\
    &=\sin\left(\frac{\delta\theta}{2}\right) \\
    &= \frac{\left(\frac{\chi}{2}\right)}{\sqrt{g_{\mathrm{bs}}^2+\left(\frac{\chi}{2}\right)^2}}.
\end{align}

\section*{ACKNOWLEDGEMENTS}
We acknowledge N. Ofek, P. Reinhold, and Y. Liu for
their work building the FPGA firmware and software, V. Sivak for providing the Josephson Array Mode Parametric Amplifier (JAMPA), and V. Joshi, G. Liu, and M. Malnou for their work on the design and assembly of the lumped element SNAIL parametric amplifiers (LSPAs) used in the experiments. The LSPAs were fabricated by the NIST Quantum Sensors Group with support from the NIST program on scalable superconducting computing. We also thank B. Brock, K. Chou, A. Koottandavida, W. Kalfus, A. Maiti, N. Mehta, A. Narla, T. Shemma and C. Zhou for useful discussions and feedback. This research was sponsored by the Army Research Office (ARO) under grant W911NF-23-1-0051, by the Air Force Office of Scientific Research (AFOSR) under grant FA9550-21-1-0209 and by the U.S. Department of Energy (DoE), Office of Science, National Quantum Information Science Research Centers, Co-design Center for Quantum Advantage (C2QA) under contract number DE-SC0012704. The views and conclusions contained in this document are those of the authors and should not be interpreted as representing the official policies, either expressed or implied, of the ARO, AFOSR, DoE or the US Government. The US Government is authorized to reproduce and distribute reprints for Government purposes notwithstanding any copyright notation herein. Fabrication facilities use was supported by the Yale Institute for Nanoscience and Quantum Engineering (YINQE) and the Yale SEAS Cleanroom.

\section*{DATA AND CODE AVAILABILITY}
The data that support the findings of this study, and the code used to perform simulations are available from the corresponding author upon reasonable request.

\section*{AUTHOR CONTRIBUTIONS}
S.J.d.G., S.H.X. and R.J.S. conceived the experiment. S.J.d.G, S.H.X. and B.J.C. set up the experiment hardware. B.J.C. fabricated the devices with help from L.F.. J.T., T.T., P.W. and J.W.O.G. advised on characterizing error sources in the dual-rail encoding and the comparison of this erasure check to prior schemes. K.M.C. and S.P. ran simulations of the surface code threshold in the presence of false negatives. S.J.d.G. and S.H.X. performed the measurements, analyzed the data, and wrote the manuscript with input from all authors.

\section*{COMPETING INTERESTS}
L.F. and R.J.S. are founders and shareholders of Quantum Circuits Inc. (QCI)

\bibliography{BeamsplitterControlPaper.bib}
\appendix

\section{Constructing a tunable CPHASE($\theta$) gate using joint-photon number-splitting}
\label{app:CPHASE}

\begin{figure*}
    \centering
    \includegraphics[width=1\linewidth]{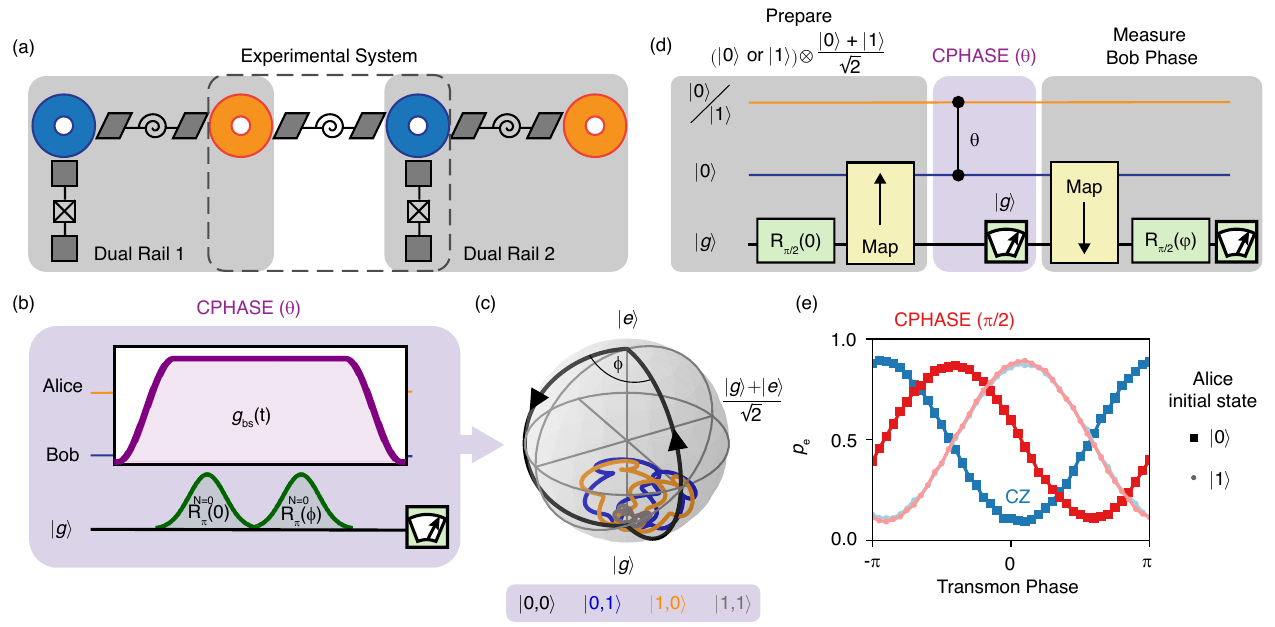}
    \caption{\textbf{CPHASE($\theta$) gate.} (a) Illustration of two adjacent dual-rail cavity qubits, with beamsplitter couplings enabled by SNAILs. The orange and blue annuli represent lambda/4 post cavities viewed from above. The setup in this experiment (within the dashed lines) can be used to operate a gate between these two hypothetical encoded dual-rail qubits. (b) Pulse sequence to obtain transmon trajectories during joint-SNAP CPHASE($\theta$) gate. (c) Bloch sphere trajectories of the transmon state conditioned on starting in every valid state for two oscillators shared between two dual-rail qubits. (d) Ramsey sequence used to probe the phase acquired by Bob's oscillator, conditioned on the state in Alice. (e) Results of Ramsey experiment to probe the phase acquired by Bob's oscillator during CPHASE($\theta$) gate with $\theta=\pi/2$ (red) and $\theta=\pi$ (blue, CZ), when Alice oscillator state is $\ket{0}$ (connected squares) and $\ket{1}$ (solid lines).}
    \label{fig:CPHASE_gate}
\end{figure*}

Just as the joint-photon-number-splitting regime enables joint-photon-number selective measurements, it can also be used to extend the established selective-number-arbitrary-phase (SNAP) gate for a single oscillator \cite{heeres_cavity_2015} to a multi-oscillator `joint-SNAP' (with some caveats). SNAP relies on the number-splitting regime to perform two consecutive photon-number-selective $\pi$-pulses with different phases, such that they enclose a geometric phase on the Bloch sphere of the transmon, thereby applying photon-number selective phases to the oscillator state. 

Similarly, the application of a strong beamsplitter drive during the same SNAP sequence makes the constituent transmon $\pi$-pulses selective on the joint photon number in two oscillators, thereby allowing one to apply an arbitrary phase to each joint-photon-number component of the two-oscillator state:
\begin{equation}
    \hat{U}_{\text{Joint-SNAP}} = \sum_{N=0}^{N_\text{max}} e^{i\phi_N} \hat{P}_N,
\end{equation}
where $\hat{P}_N$ is the projector onto two-oscillator states with total photon number $N$ and $N_{\text{max}}$ is the highest total photon number state we wish to address. However, as is described in App.~\ref{app:large_N}, for $N_{\text{max}} > 1$, an increasingly large ratio $|g_{\text{bs}}/\chi|$ is required to ensure that different states with the same $N$ evolve together. This construction is therefore most useful for applying a phase conditioned on $N=0$ total photons.

The application of an $N=0$\ -\ selective phase implements a tunable CPHASE($\theta$) gate on two Fock-encoded qubits or equivalently, if the two oscillators are each part of a different dual-rail qubit (as in Fig.~\ref{fig:CPHASE_gate}a), a logical CPHASE($\theta$) gate on dual-rail qubits. This gate is a maximally entangling gate between qubits, a key ingredient for the dual-rail architecture, and consists of applying a tunable phase only to the $\ket{1_\text{L}}\ket{1_\text{L}}$ two-qubit dual-rail state while leaving other dual-rail states untouched. Since the logical information for a single dual-rail qubit is redundantly encoded in two oscillators, we can effect this by applying a phase to the $\ket{0,0}$ Fock state of the two addressed oscillators. The application of a geometric phase to the joint-vacuum state of two oscillators has previously been shown in the context of Schrodinger cat qubits using a transmon with a static dispersive coupling to both modes \cite{xu_demonstration_2020} but is here achieved with a transmon statically coupled to one mode and a strong beamsplitter interaction that temporarily activates the coupling to the other. (In principle, applying the phase to any \textit{one} of $\ket{0,0}$, $\ket{0,0}$, $\ket{0,0}$ or $\ket{0,0}$, combined with local rotations on the cavities, can effect a CPHASE gate.)

The constraints on the value of $|g_{\mathrm{bs}}/\chi|$ are slightly different in the case of the CPHASE($\theta$) gate than for the erasure check. Whereas previously we only needed to avoid transitions in the $N=1$ total photon number manifold, the $\ket{1,1}$ Fock state (with $N=2$) is a valid initial state for two oscillators belonging to two different dual-rail qubits. As such, the need to drive an $N=0$ transition while avoiding both $N=1$ and $N=2$ transitions leads us to choose a value of $|g_{\mathrm{bs}}/\chi| \approx 1.6$ (see triangle Fig.~\ref{fig:driven_spec}c). Further increasing $|g_{\text{bs}}/\chi|$ would increase the separation to unwanted transitions at the potential cost of additional decoherence as we approach the maximum value accessible in this device.

The simulated Bloch sphere trajectories of the transmon state for all valid input oscillator states (shown in Fig.~\ref{fig:CPHASE_gate}c) show the principle of this operation for CPHASE($\pi$) = CZ. As shown in Fig.~\ref{fig:CPHASE_gate}b, $\ket{0,0}, \ket{0,1}, \ket{1,0}$ and $\ket{1,1}$ are prepared, before performing back-to-back selective $\pi$-pulses with a phase difference that is chosen to ensure that $\theta = \pi$. Whereas for initial states $\ket{0,1}, \ket{1,0}$ and $\ket{1,1}$ the transmon trajectories stay near the ground state and return to $\ket{g}$, the $\ket{0,0}$ trajectory passes through $\ket{e}$ and encloses a large geometric phase before returning to $\ket{g}$.

To verify that this protocol performs a  CPHASE($\theta$) gate on two Fock-encoded qubits, we can perform the Ramsey sequence shown in Fig.~\ref{fig:CPHASE_gate}d. This starts by preparing either $\ket{0}\ket{+}$ or $\ket{1}\ket{+}$ in the oscillators, where $\ket{+} = (\ket{0}+\ket{1})/\sqrt{2}$. The superposition states are loaded into the oscillator by preparing $(\ket{g}+\ket{e})/\sqrt{2}$ in the transmon and performing an `encoding' optimal control pulse (OCP) which maps $\ket{\psi}_{\text{qubit}}\ket{0}_{\text{oscillator}}\rightarrow\ket{g}_{\text{qubit}}\ket{\psi}_{\text{oscillator}}$. We then enact the CPHASE($\theta$) gate for variable $\theta$, including a transmon readout to catch errors that leave it out of the ground state, before mapping the oscillator state onto the transmon with a `decoding' OCP (which performs the inverse of the `encoding' OCP) and measuring the phase of the transmon using a variable-phase $\pi/2$-pulse. In this way, the transmon measurement allows us to probe the phase acquired by the Bob oscillator state during the gate.

The results for $\theta=\pi/2$ and $\theta=\pi$, shown in Fig.~\ref{fig:CPHASE_gate}e, show that a tunable phase can be imparted on Bob's superposition state dependent on the state in Alice's cavity. The measured transmon population oscillations encode the phase of Bob's cavity superposition. When Alice is initialized in $\ket{1}$ (connected dots) these oscillations lie on top of one another, indicating that the two-oscillator states $\ket{1,0}$ and $\ket{1,1}$ have the same relative phase regardless of the choice of $\theta$. Meanwhile the relative position of the oscillations when Alice is initialized in $\ket{0}$ (squares), can be tuned depending on the enclosed phase, such that we are free to choose the relative phase between the two-oscillator states $\ket{0,0}$ and $\ket{0,1}$. Together with arbitrary rotations on either cavity, which can be performed virtually, this provides the necessary control for a CPHASE($\theta$) gate. The loss in contrast of the oscillations is dominated by SPAM errors during the OCT encode and decode pulses, as well as cavity photon loss. Full characterization of the dual-rail CPHASE gate performance would require process tomography and would ideally be performed on a system with four modes comprising two dual-rails (as shown in Fig.~\ref{fig:CPHASE_gate}a). This would allow for both high-fidelity SPAM and the detection of photon loss during the gate.

\begin{figure}
    \centering
    \includegraphics[width=1\linewidth]{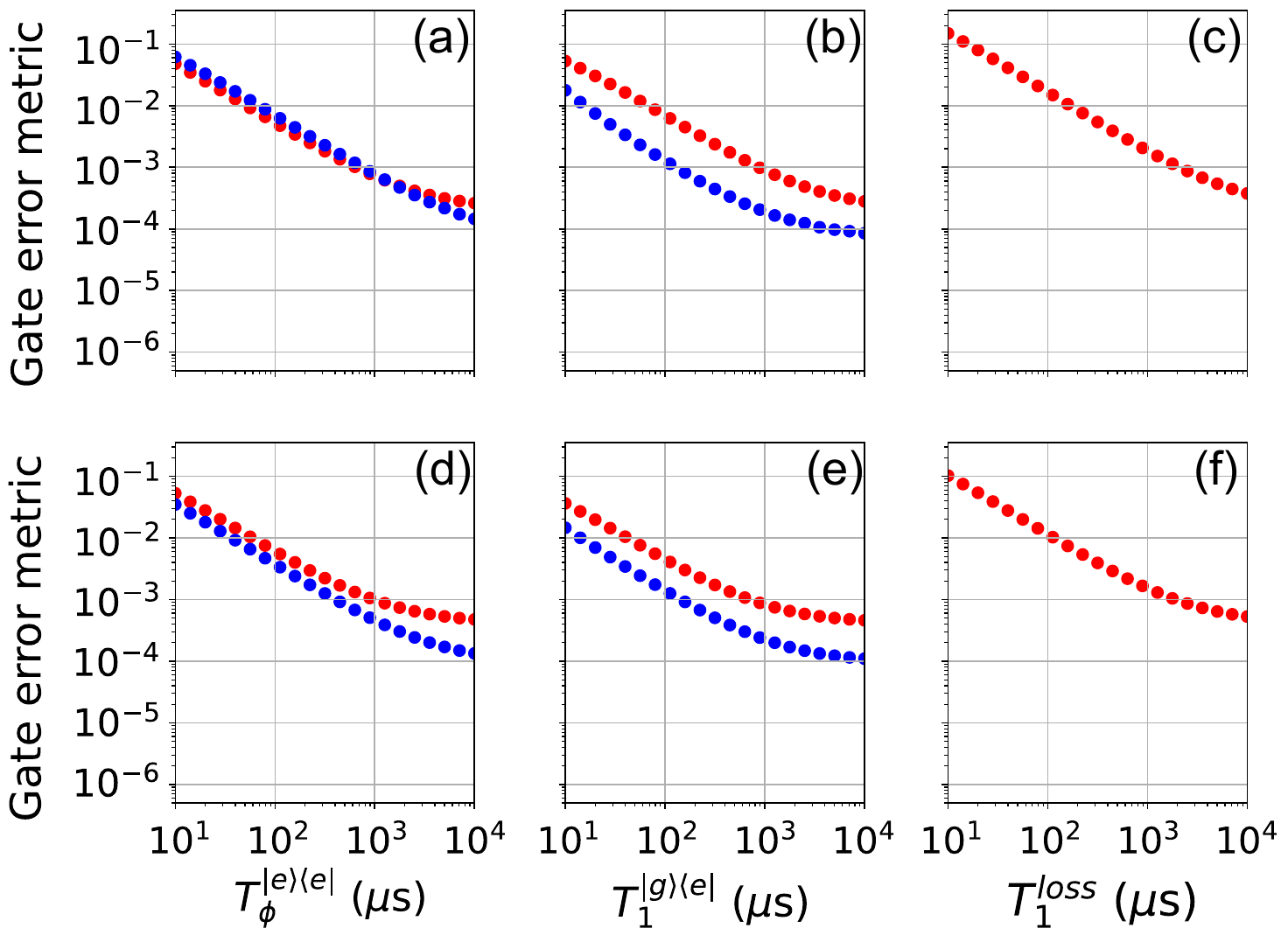}
    \caption{\textbf{Projected dual-rail CPHASE performance.} Simulated erasure rate (red) and post-selected infidelity (blue) for the dual-rail CPHASE($\pi$) gate, with chopped Gaussian transmon pulse of total duration $3.194~\muup$s (a-c) and `shaped-square' transmon pulse shape of total duration $1.994~\muup$s (d-f). In each column, either the transmon dephasing rate $T_\phi^{\ket{e}\bra{e}}$, transmon decay rate $T_1^{\ket{g}\bra{e}}$ or cavity photon loss rates $T_1^{\text{loss}}$ are varied with no other loss channels included. All simulation results are averaged over all 36 two-qubit states formed from every combination of the 6 dual-rail cardinal states in either qubit. For all simulations, we use a dispersive shift of $\chi/2\pi = 1$~MHz, and an infinite-bandwidth square beamsplitter pulse whose amplitude is optimized to minimize coherent errors while restricting it to obey $g_{\mathrm{bs}}/2\pi < 2~$MHz. Impact of Stark shifts due to the beamsplitter drive is neglected.}
    \label{fig:CPHASE_scaling}
\end{figure}

This gate construction has some, but not all of the desired features for a multi-cavity encoding scheme such as the dual-rail. By appending the CPHASE($\theta$) operation with a transmon readout, some, but not all, transmon errors can be caught and converted to erasures. Fig.~\ref{fig:CPHASE_scaling} shows the simulated erasure rate and post-selected state infidelity, which depend linearly on the underlying physical error rates but with different offsets. (The plateau at long coherence times is due to residual coherent errors that could be removed by using a pulse shape with more degrees of freedom than a simple square pulse.) Whereas the post-selected infidelity due to transmon decay errors is distinguishably lower than the corresponding erasure rate, indicating that most transmon decay errors can be caught by the check, for transmon dephasing errors they are comparable.

This construction therefore does not achieve the full error-detecting and fault-tolerant properties of the $ZZ(\theta)$ gate for dual-rail qubits proposed by Tsunoda et al. \cite{tsunoda_error-detectable_2023}. Adapting this CPHASE($\theta$) gate to be fully error-detectable to transmon errors would require the use of three transmon energy levels (to protect against transmon relaxation) and simultaneous transmon drives on all joint-photon-number-selective peaks (to protect against transmon dephasing) \cite{reinhold_error-corrected_2020}.

\section{Calibrating beamsplitter strength and Stark shifts}
\label{app:calibration}

\begin{figure}
    \centering
    \includegraphics[width=1\linewidth]{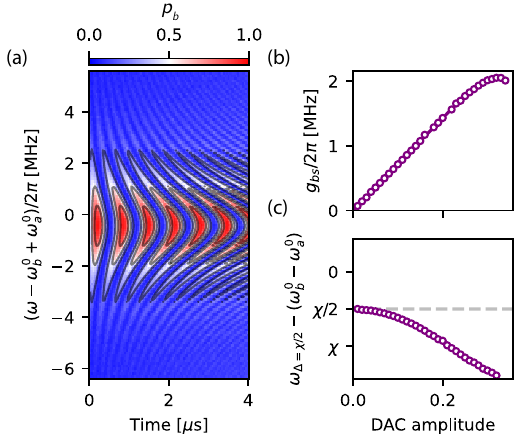}
    \caption{\textbf{Beamsplitter amplitude and resonance frequency.} (a) Measured oscillations in the population of Bob's cavity, $p_b$, as a function of applied beamsplitter drive frequency. The data is shown for a DAC amplitude of 0.25 for the beamsplitter drive. Contours show the fit to Eq.~\ref{eq:chevron_fit}, from which $g_{\text{bs}}$ and a first estimate for the resonance frequency are obtained. (b) Fitted beamsplitter amplitude $g_{\text{bs}}$ as a function of DAC amplitude. (c) Beamsplitter drive frequency required to ensure a symmetric detuning $\Delta=\chi/2$ as a function of DAC amplitude, relative to the bare resonance frequency.}
    \label{fig:gbs_stark_cal}
\end{figure}

The beamsplitter amplitude $g_{bs}$ and resonance frequency $\omega_{\Delta=0}$ (i.e. the drive frequency required to achieve a detuning $\Delta=0$) are measured by initializing the state $\ket{0,1}\otimes\ket{g}$, turning on the beamsplitter drive and measuring the photon number in Bob's cavity after some time $t$. By varying the frequency and duration of the beamsplitter drive, we observe the chevron pattern seen in Fig.~\ref{fig:gbs_stark_cal}a, which we can fit to:
\begin{align}
    P_1(\omega, t) = A \bigg[ &\cos^2\left(\frac{\Omega t}{2}+\phi\right) \nonumber \\ + &\left(\frac{\omega-\omega_{\Delta=0}}{\Omega}\right)^2\sin^2\left(\frac{\Omega t}{2} + \phi\right)\bigg]  + c,
    \label{eq:chevron_fit}
\end{align}
where the detuned oscillation rate is given by
\begin{equation}
    \Omega = \sqrt{g_{\mathrm{bs}}^2+\left(\omega - \omega_{\Delta=0}\right)^2},
\end{equation}
$\phi$ captures the phase accumulation while the beamsplitter drive is ramping up, and $A$ and $c$ capture the contrast and offset due to SPAM.

Fig.~\ref{fig:gbs_stark_cal}b shows the fitted values of $g_{\mathrm{bs}}$ over the range of microwave drive amplitudes used in this experiment. As described in previous work \cite{chapman_high--off-ratio_2023}, the measured beamsplitter amplitude is linearly  proportional to drive amplitude at low amplitudes but deviates at high amplitudes. In order to interpolate between the amplitudes used for the calibration experiment, we fit the $g_{\mathrm{bs}}$ curve to a degree-5 polynomial (with the intercept forced to pass through the origin).

The fits to the chevrons also provide an estimate of the beamsplitter resonance frequency, which at low pump powers is equal to the difference in the undriven cavity frequencies, $\omega_b^0 - \omega_a^0$. This in turn allows us to estimate the frequency $\omega_{\Delta=\chi/2}$ at which to drive the beamsplitter to achieve a `symmetric' detuning $\Delta = \chi/2$. This frequency shifts downwards due to Stark shifts of the cavity frequencies induced by the beamsplitter drive, with Bob's cavity frequency shifting downward by more than Alice's:
\begin{align}
\omega_{\Delta=\chi/2} = &~\omega_{\Delta=0} + \chi/2 \nonumber\\ = &\left(\omega_b^0+ \delta^{\text{Stark}}_b\right)-\left(\omega_a^0  + \delta^{\text{Stark}}_a\right) + \chi/2 \nonumber\\
=&\left(\omega_b^0-\omega_a^0+\chi/2\right) + \nonumber \\
&\quad\left(\delta^{\text{Stark}}_b - \delta^{\text{Stark}}_a\right).
\end{align}

We perform a finer calibration of this frequency by measuring the oscillations of the cavities' photon number induced by the beamsplitter drive when the transmon is in $\ket{g}$ or in $\ket{e}$. The point at which the frequencies of these oscillations are matched marks the symmetric $\Delta = \chi/2$ point for which the data in Fig.~\ref{fig:driven_spec} is taken. These values, relative to $\omega_b^0 - \omega_a^0$,  are plotted in Fig.~\ref{fig:gbs_stark_cal}c.

\section{Measuring transition matrix elements}
\label{app:matrix_elements}
To measure the transition matrix elements in Fig.~\ref{fig:energy_levels_matrix_elements}b,  we perform a power-Rabi experiment which determines the amplitude of the spectroscopy tone required to perform a $\pi$-pulse on the ancilla. This quantity is inversely proportional to the transition matrix element. We run this sequence at each value of $|g_{\mathrm{bs}}/\chi|$ for both the central $\delta m=0$ transition and the lower $\delta m =-1$ transition. The $\delta m = -1$ transition is chosen over the $\delta m = +1$ transition, owing to its distance from other transitions, allowing for a cleaner measurement, but both $\delta m=\pm1$ transitions are expected to share the same matrix element.

\begin{figure}
    \centering
    \includegraphics[width=1\linewidth]{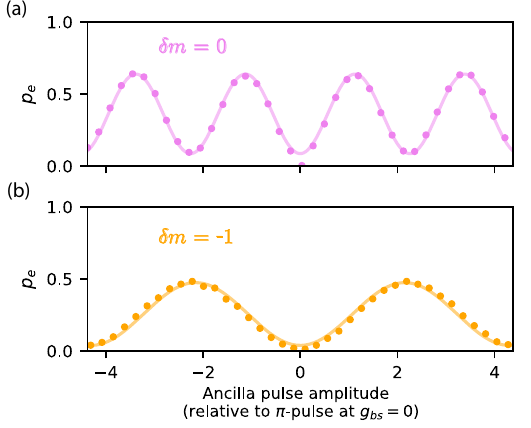}
    \caption{\textbf{Power Rabi oscillations used to extract matrix elements.} Measured probability of finding the ancilla in $\ket{e}$ after a selective ancilla pulse of variable amplitude for the $\delta m = 0$ (a) and $\delta m = -1$ (b) transitions in the $N=1$ photon manifold of Fig.~\ref{fig:driven_spec}c. A symmetric beamsplitter drive is applied during the pulse, in this case with an amplitude $g_{\text{bs}}\approx|\chi|$. The ancilla pulse amplitude is normalized to the value required for a $\pi$-pulse in the absence of a beamsplitter drive. The frequency of these oscillations is used to extract the transition matrix elements in Fig.~\ref{fig:energy_levels_matrix_elements}.}
    \label{fig:rabi_cal}
\end{figure}

This measurement requires us to ensure that the frequency of the spectroscopy tone is on resonance with the transition. If the spectroscopy tone has some detuning, the rate of Rabi oscillations would be overestimated. To mitigate this, we precisely find the ancilla resonance frequency for each value of $|g_{\mathrm{bs}}/\chi|$ by taking a vertical slice of the $N=1$ colorplot in Fig.~\ref{fig:driven_spec}c in the vicinity of the desired transition, fitting it to a Gaussian and extracting the fitted center frequency.

The power-Rabi sequence consists of initializing the oscillators in the state $(\ket{0,1}+i\ket{1,0})/\sqrt{2}$, playing a Gaussian-shaped pulse of total duration $14.8~\muup s$ with varying amplitude and measuring the state of the transmon at the end. The rate of the oscillations (see Fig.~\ref{fig:rabi_cal} for an example) is inversely proportional to the amplitude required for a single $\pi$-pulse, and so is proportional to the transition matrix element.
We avoid the need to precisely calibrate the drive power delivered to the package by normalizing all the power-Rabi oscillation frequencies to the value at $g_{\mathrm{bs}}=0$ for $\delta m =-1$. Importantly, this technique is not sensitive to infidelity in state preparation and measurement, as this just affects the amplitude of the oscillations, not their rate.
The experimental $\delta m = 0$ curve in Fig.~\ref{fig:energy_levels_matrix_elements}b does not extend to $g_{\mathrm{bs}}/\chi = 0$. The reason for this is that as the transition matrix element decreases, the transmon drive amplitude $|\epsilon|$ required to perform a $\pi$-pulse increases. As $|\epsilon|$ becomes comparable to $\max(g_{\mathrm{bs}}, \chi)$, the assumption that the transmon drive can be treated as a perturbation in the Hamiltonian breaks down. Specifically, we find that there is no longer a single degenerate $\delta m = 0$ transition at $\delta\omega = 0$.

\section{Degeneracy breaking for non-symmetric beamsplitter detuning}
\label{app:degeneracy_breaking}

\begin{figure}
    \centering
    \includegraphics[width=1\linewidth]{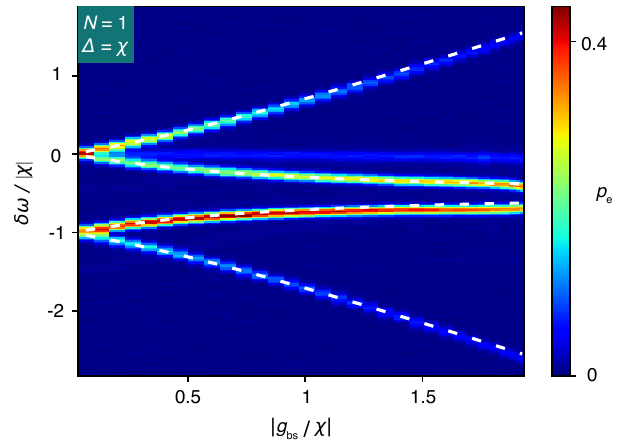}
    \caption{\textbf{Driven transmon spectrum with non-symmetric beamsplitter detuning.} Measured probability of exciting the transmon when initializing the cavities in $\ket{\psi} = (\ket{0,1} + i\ket{1,0})/\sqrt{2}$ and applying a selective transmon pulse in the presence of a beamsplitter drive with amplitude $g_{\mathrm{bs}}$ and detuning $\Delta=\chi$. The white dashed lines show predicted transition frequencies from spin-oscillator model.}
    \label{fig:non_sym}
\end{figure}

In the main text, we primarily consider the case where a beamsplitter drive is applied with a symmetric detuning, $\Delta = \chi/2$, so-called because it lies exactly halfway between the beamsplitter resonance when the ancilla qubit is in $\ket{g}$ and the resonance when the ancilla is in $\ket{e}$. In this case, many of the transition frequencies are degenerate. However, as can be seen in Eq.~\ref{eqn:nondegen_freqs}, we expect this degeneracy to break when we move to a non-symmetric beamsplitter detuning, $\Delta \neq \chi/2$. For example, in the case that the total photon number $N = 1$, we expect four (rather than three) separate transition frequencies.

We can verify this by measuring the transmon spectrum when the cavities are initialized in $\ket{\psi} = (\ket{0,1} + i\ket{1,0})/\sqrt{2}$  and we apply a beamsplitter drive with detuning $\Delta = \chi$, shown in Fig.~\ref{fig:non_sym}. The white dashed lines show the theory predictions which agree well with the measured spectrum, with the exception of a slight shift downwards in frequency of the experimental curves with respect to theory at large beamsplitter amplitudes. This is at least in part due to a slight negative Stark shift of the transmon frequency when a strong beamsplitter drive is applied.
\section{Tune-up of joint-photon number-selective pulses}
\label{app:tuneup}

In general, a variety of transmon pulse shapes may be chosen for the joint-photon number-selective pulse. The choice of pulse shape provides a way to trade off between idling and false negative errors (reduced with a shorter, less frequency selective pulse) and false positive and induced Pauli errors (reduced with a longer, more frequency selective pulse). In the erasure check demonstrated in this work, we considered a cosine-ramped square pulse. Later in this section we also provide a method for tuning up a longer chopped Gaussian pulse, as used for the CPHASE gate in Appendix~\ref{app:CPHASE}.

\subsubsection*{Erasure check with square pulse}
We first consider cosine-ramped square transmon pulses of the form
\begin{equation}
    \frac{\hat{\mathcal{H}}_{\text{transmon}}}{\hbar} = f(t)\ \hat{\sigma}_x,
\end{equation}
with
\begin{equation}
    f(t) =
    \begin{cases}
    \frac{A}{2}\left(1-\cos\left(\frac{\pi t}{t_{r}}\right)\right), & 0<t<t_{r} \\
    A, & t_{r} \leq t \leq T_\mathrm{p}-t_r \\
    \frac{A}{2}\left(1+\cos\left(\frac{\pi \left(t-t_{r}+T_\mathrm{p}\right)}{t_{r}}\right)\right), & T_\mathrm{p}-t_r<t<T_\mathrm{p}.
    \end{cases}
\end{equation}
where $A$ is the pulse amplitude, $T_p$ is the pulse duration and $t_r$ is the duration of the cosine ramp. The ramp time $t_r$ should be kept sufficiently long to prevent coherent errors from the $\ket{f}$ level of the transmon. 

When using a square transmon pulse, there are discrete operating points that ensure that dual-rail and ancilla states both return to where they began. As mentioned in the main text, we find from simulation that aligning the center (in time) of the cosine ramps of the transmon pulse with those of the beamsplitter drive ensures optimal performance.

\begin{figure}
    \centering
    \includegraphics[width=1\linewidth]{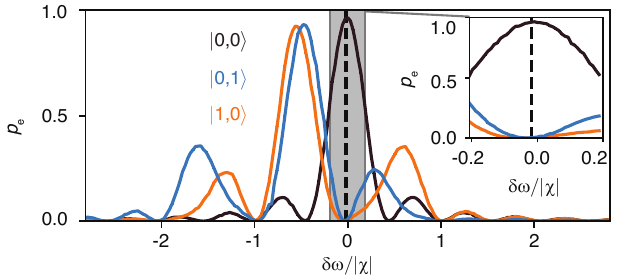}
    \caption{\textbf{Square Pulse Spectroscopy.} Probability of erasure check reading out $\ket{e}$ as a function of transmon pulse detuning during the check when initializing the dual-rail qubit in $\ket{0,0}$ (black), $\ket{0,1}$ (blue) or $\ket{1,0}$ (orange) state. The results are post-selected on the total photon number in the oscillators remaining the same after the check.}
    \label{fig:square_pulse_spec}
\end{figure}

The tune-up procedure consists of:
\begin{enumerate}
    \item Calibrate the following quantities as a function of DAC amplitude:
    \begin{itemize}
        \item beamsplitter amplitude $g_{bs}$,
        \item beamsplitter detuning $\omega_{\Delta=\chi/2}$,
        \item ancilla resonance frequency $\omega_{\text{ancilla}}$.
    \end{itemize}
    As described in Appendix~\ref{app:calibration}, the frequencies may shift as the drive amplitude is increased.

    \item Identify the maximum beamsplitter amplitude $g_{\text{bs}}^{\text{max}}$ that can be accessed without introducing significantly more errors than when idling. Above certain drive amplitudes, one may see enhanced photon loss, dephasing or heating.
    
    \item Find an initial starting point for the operating parameters: $T_p$, $g_{\text{bs}}$, $A$, the beamsplitter drive frequency $\omega$, and the detuning of the ancilla drive from its undriven resonance frequency $\delta\omega$. For infinite bandwidth pulses ($t_r=0$) and no Stark shifts, the following analytical expressions approximate the ideal values:
    \begin{align}
        T_p &= \frac{2\pi\sqrt{3}}{|\chi|} & g_{\text{bs}} &= |\chi|\sqrt{\frac{m^2}{3}-\frac{1}{4}} \\
        A &= \frac{|\chi|}{4\sqrt{3}} & \omega &= \omega_b - \omega_a + \chi/2,
    \end{align}
    with $\delta\omega=0$ and where the integer $m>1$ can be varied in order to operate at different values of $g_{\text{bs}}$. We choose values close to $m=2$ to ensure $g_{\text{bs}} < g_{\text{bs}}^{\text{max}}$.
    
    A more precise starting point can be obtained by performing a Schr\"odinger equation simulation which includes the ramp times as well as Stark shifts (i.e. how the ancilla resonance frequency and $\omega_{\Delta=\chi/2}$ change with pump amplitude) and running gradient descent optimization using the infidelity of the state transfers
    \begin{align}
        \ket{0,0,g} &\rightarrow \ket{0,0,e}, \\
        \ket{0,1,g} &\rightarrow \ket{0,1,g}, \\
        \ket{1,0,g} &\rightarrow \ket{1,0,g}
    \end{align}
    as a cost function.

    \item Starting with the parameters found above, fine-tune the ancilla drive amplitude $A$ and detuning $\delta\omega$ experimentally by initializing the cavities in $\ket{0,0}$ and performing the ancilla pulse with the beamsplitter drive applied. Choose the parameters that maximize the probability of exciting the transmon, thereby minimizing false negatives. Since the $\ket{0,0}$ state is unaffected by the beamsplitter drive, this calibration depends very weakly on changes in $g_{\text{bs}}$ in the following step.

    \item With $T_p$ and $A$ fixed, perform the following experiment while sweeping the values of $g_{\text{bs}}$ and $\omega$:
    \begin{itemize}
        \item Initialize the system in $\ket{0,1,g}$ or $\ket{1,0,g}$,
        \item Perform $N$ successive erasure checks,
        \item Perform an erasure-detected logical measurement.
    \end{itemize}
    From this single experiment, we obtain three key metrics to maximize:
    \begin{itemize}
        \item Probability of passing $N$ checks when initializing in $\ket{0,1,g}$,
        \item Probability of passing $N$ checks when initializing in $\ket{1,0,g}$,
        \item Probability of returning to $\ket{0,1,g}$ when initializing in $\ket{0,1,g}$.
    \end{itemize}
    With the optimal choice of $T_p$, there should be a point in $(g_{\text{bs}}, \omega)$ space that maximizes all three, thereby minimizing both false positives and coherent Pauli errors. 
    
    \item If there is not optimal choice for $g_{\text{bs}}$ and $ \omega$, $T_p$ can be adjusted. If the operating points that optimized for false positives lie at a lower $g_{\text{bs}}$ than is required to ensure the dual-rail cavity states return to where they began, the pulse duration $T_p$ can be increased (and vice-versa). The previous step can then be repeated. 

    We can probe whether the pulse duration is set correctly via a spectroscopy experiment. This involves preparing the cavities in $\ket{0,0}$, $\ket{0,1}$ and $\ket{1,0}$, and performing a single erasure check with a variable detuning $\delta\omega$ on the ancilla pulse. We also post-select on the total photon number in the two cavities remaining the same after the check to remove the effect of photon loss or gain. Fig.~\ref{fig:square_pulse_spec} shows the result for the parameters optimized in this experiment, with the maximum of the $\ket{0,0}$ peak lining up with the minimum of the $\ket{0,1}$ and $\ket{1,0}$ traces. The frequency at which they line up (indicated with the dashed vertical line) is the ancilla frequency detuning used for the check. The slight gap between the peaks of the $\ket{0,1}$ and $\ket{1,0}$ curves is expected, even at $\Delta=\chi/2$, and is due to the ancilla drive amplitude no longer being much smaller than $g_{\text{bs}}$. If the extrema are not aligned at zero detuning, increasing (decreasing) the duration of the square pulse allows us to narrow (widen) the spectrum of the curves such that they align. Doing so will then require re-optimizing $g_{\mathrm{bs}}$ and $\omega$ in this iterative tune-up scheme.

    \item Finally, double-check the calibration of $A$ and $\delta\omega$ now that $g_{\text{bs}}$ and $\omega$ have been adjusted.
    
\end{enumerate}

\subsubsection*{Erasure Check with Gaussian pulse}
Alternatively, one can consider (amongst other narrow bandwidth pulse shapes) chopped-Gaussian pulses of the form:
\begin{equation}
    f(t) = A\left[\exp\left(-\frac{\left(t-n_{\text{chop}}\sigma\right)^2}{2\sigma^2}\right) - \exp\left(-\frac{n_{\text{chop}}^2}{2}\right)\right],
\end{equation}
where $\sigma$ is the RMS width of the Gaussian and the pulse duration is $T_\mathrm{p} = 2n_{\text{chop}}\sigma$.
\begin{enumerate}
    \item[1-2.] Same as for the square pulse sequence. 
    \item[3.] Set the beamsplitter amplitude to a value in the range $\sqrt{3}|\chi|/2 < g_{\text{bs}} < g_{\text{bs}}^{\text{max}}$. Larger values within this range will suppress the impact of ancilla decoherence on dual-rail Pauli errors. Set the beamsplitter frequency to $\omega_{\Delta=\chi/2}$. Tune up a $\pi$-pulse on the ancilla with the cavities in $\ket{0,0}$ for these beamsplitter parameters.
    \item[4.] Prepare $\ket{0,1}$ and $\ket{1,0}$ in the cavities and sweep the selectivity (given by $\sigma$) of the ancilla pulse while keeping the product of $A$ and $\sigma$ fixed. By observing the probability of exciting the ancilla, we can determine a minimum selectivity in order to ensure we do not flag $\ket{0,1}$ and $\ket{1,0}$ as erasures. Set $\sigma$ to a value above this threshold.
    \item[5.] Finally, we must ensure that the cavity states return to their initial states at the end of the sequence (to avoid coherent Pauli errors). Prepare $\ket{1,0}$ and $\ket{0,1}$, perform the erasure check with a variable beamsplitter amplitude, and measure the dual-rail state at the end. Depending on how close we are to performing a full revolution, there are a few options:
    \begin{itemize}
        \item If the beamsplitter amplitude can be increased to perform an integer number of revolutions without exceeding $g_{\mathrm{bs}}^{\text{max}}$, do this. This ensures the pulse is as short as it can be.
        \item If not, increase $\sigma$ until an integer number of revolutions is performed (at the same $g_{\mathrm{bs}}$).
    \end{itemize}
    \item[6.] Having made this adjustment, other parameters may need to be changed: \begin{itemize}
        \item If $g_{\mathrm{bs}}$ has been adjusted, also adjust the beamsplitter detuning (to ensure $\Delta=\chi/2$), and the transmon detuning (if it is Stark shifted by the drive).
        \item If $\sigma$ has been adjusted, fine-tune the amplitude of the transmon pulse such that a complete $\pi$-pulse is performed.
    \end{itemize} 
    \item[7.] Return to Step 5 to verify whether the dual-rail computational states return to their initial states. If not, iterate over Steps 5 and 6 until this is true.
\end{enumerate}

\subsubsection*{Adapting tune-up procedure for CPHASE($\theta$)}
When performing the CPHASE($\theta$) gate in Appendix~\ref{app:CPHASE} we also care about the input two-cavity state $\ket{1,1}$. In this experiment we adapt the Gaussian pulse tune-up procedure by (a) ensuring the starting value of $g_{\mathrm{bs}}$ is sufficiently large to avoid driving transitions in the $N=2$ manifold, and (b) adding the state $\ket{1,1}$ to the $\sigma$ and $g_{\mathrm{bs}}$ calibration sequences in steps 4 and 5.

The square pulse tune-up procedure may also be adapted for use with CPHASE($\theta$) but more degrees of freedom in the pulse shape will be required to fully avoid driving both the $N=1$ and $N=2$ peaks in the transmon spectrum.

\section{Fault-tolerant threshold simulations}
\label{app:thresholdsimulations}

We find fault-tolerant thresholds of the CSS surface code, assuming noisy two-qubit gates and mid-circuit erasure checks applied to every qubit after each gate. These noisy two-qubit gates have error probability $p$ in which its qubits experience an erasure or Pauli error. We define a parameter called the erasure fraction, $R_\mathrm{e} = p_{\mathrm{eras}}/ (p_{\mathrm{eras}} +p_\mathrm{Pauli})$. Although we anticipate an erasure fraction of $R_\mathrm{e}=0.9$, extracted from the erasure check demonstrated in this work, we do not expect $R_\mathrm{e}$ to change significantly for two-qubit gates. Namely, the beamsplitter and dispersive interactions required for the erasure checks are also utilized in proposals for entangling gates~\cite{teoh_dual-rail_2023, tsunoda_error-detectable_2023}.

With probability $p (1-R_\mathrm{e})$, both qubits involved in the gate experience a Pauli error uniformly drawn from the set $\{I,X,Y,Z\}^{ \otimes 2 }$. Furthermore, with probability $p R_\mathrm{e}$, one qubit from the two-qubit gate leaves the computational space. To describe what happens to other unleaked qubits, we assume that a leaked qubit induces computational Pauli errors on any qubit it interacts with~\cite{fowler2013coping,ghosh2013understanding,suchara2015leakage,brown2018comparing,brown2019leakage}. Thus, the unleaked qubit in the gate undergoes a Pauli error that is dependent on which two-qubit gate is applied~\cite{brown2019leakage}. Specifically, in a CZ gate, the unleaked qubit experiences an $I$ or $Z$ error with equal probability. If the control (target) qubit is leaked in a CX gate, then the unleaked target (control) qubit experiences an $I$ or $X$ ($I$ or $Z$) error.
The proposed method for two-qubit gates in the dual-rail cavity architecture, the $ZZ(\theta)$ gate, follows the specific model of leakage-induced Pauli errors studied here~\cite{teoh_dual-rail_2023, tsunoda_error-detectable_2023}. For this gate proposal, one can show that a leaked qubit induces rotational errors on the other qubit along one Pauli axis.

We model false negatives of erasure checks as the following. When erasure occurs during a two-qubit gate, the erasure check after this gate returns no flag with the false-negative probability $3.7\%$. Given a false negative, we assume the detectable leakage will be detected during the next round of mid-circuit erasure checks. We therefore ignore the effect of doubly-missed erasures, which are expected to occur with a probability on the order of $10^{-5}$. These events, while damaging in the context of error-correction and a subject for future work, are rare relative to leakage rates in state-of-the-art transmon-based architectures (on the order of $10^{-3}$) \cite{miao2023overcoming}. As before, the leaked qubit is assumed to induce computational errors on any qubit it interacts with before it is detected. To account for these correlated errors, we use a modified, weighted union-find decoder. For details on this decoder, we refer to~\cite{chang2024erasure}.

\section{Erasure-detected end-of-line logical measurement}
\label{app:EOL}

\begin{figure*}
    \centering
    \includegraphics[width=1\linewidth]{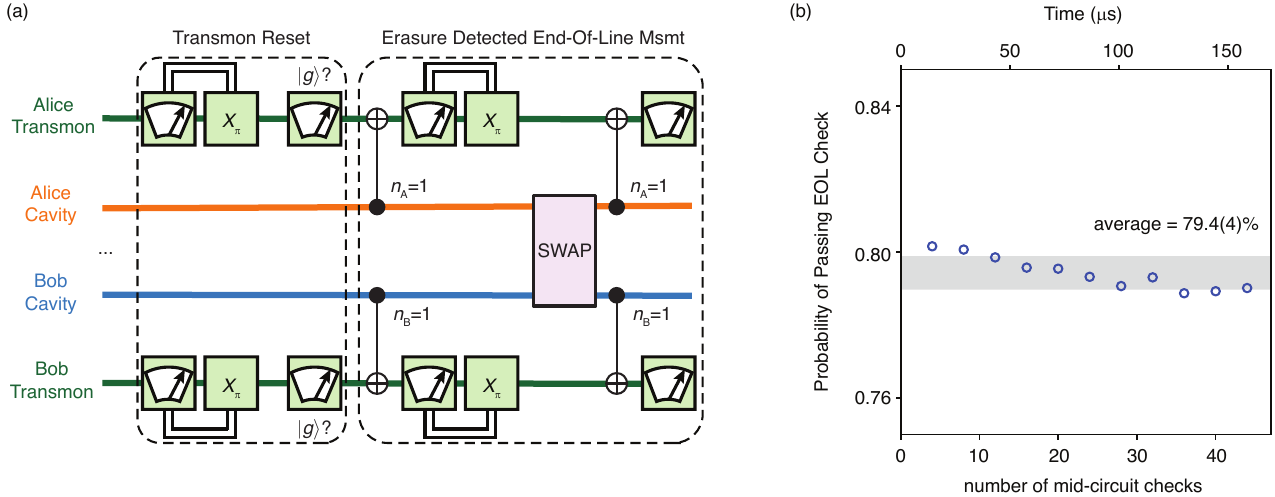}
    \caption{\textbf{Erasure-detected end-of-line (EOL) logical measurement.} (a) Pulse sequence for the erasure detected EOL check comprises transmon reset and erasure-detected EOL measurement.(b) Probability of passing the EOL measurement plotted against the number of preceding mid-circuit erasure checks before the EOL measurement. Gray horizontal line shows average probability.}
    \label{fig:EOL}
\end{figure*}

Logical X, Y and Z measurements at the end of the sequence in Fig.~\ref{fig:Erasure_Check}d consist of a 50:50 beamsplitter pulse that rotates the dual-rail state onto the desired measurement axis, followed by an erasure-detected end-of-line (EOL) measurement, illustrated in Fig.~\ref{fig:EOL}a \cite{chou_demonstrating_2023}. In this section, we describe the measurement sequence we use and show that the post-selection probability does not depend on the number of preceding successful mid-circuit checks.

The logical measurement starts with a transmon reset sequence that conditionally resets both transmons and double checks that they are indeed in $\ket{g}$ after the reset. Resetting the transmons to $\ket{g}$ is crucial because starting the EOL measurement with one or more transmon in $\ket{e}$ could lead to an error in state assignment.  

Subsequently, cavity-photon-number-selective $\pi$ pulses excite the measurement transmons conditioned on $1$ photon in the adjacent cavity. We record this result and conditionally reset the transmons to $\ket{g}$ before swapping the cavity states with two consecutive $50:50$ beamsplitter pulses. We then perform a second photon-number-selective measurement. This echoed sequence, where the state of each cavity is measured twice (once with each measurement transmon), removes bias due to any differences between the measurement fidelities of the transmons or between the coherences of the cavities. We only keep results in which (a) these two rounds of measurement are consistent and (b) they indicate a total of 1 photon in both cavities (i.e. $\ket{g}_A\ket{e}_B$ followed by $\ket{e}_A\ket{g}_B$, or $\ket{e}_A\ket{g}_B$ followed by $\ket{g}_A\ket{e}_B$).

Fig.~\ref{fig:EOL}b shows the probability of passing the EOL measurement, conditioned on passing all previous mid-circuit erasure checks, for the data presented in Fig.~\ref{fig:Erasure_Check}f. This probability is consistent regardless of the number of mid-circuit checks performed thus confirming that the EOL measurement does not introduce any bias in our measurements. This `strict' EOL measurement sequence discards a relatively large fraction of the data, with an average success probability of $79.4 \pm 0.4\%$, but in doing so, ensures a faithful measurement of the final dual-rail states.

\section{Dual-rail state fidelity after repeated erasure checks}
\label{app:QPT}
To characterize the Pauli error rate per erasure check (post-selected on passing every check), we perform the sequence shown in Fig.~\ref{fig:Erasure_Check}d, in which we prepare all six dual-rail cardinal states, perform $n$ successive erasure checks (with an echo pulse inserted halfway) and then perform a logical measurement along the axis in which the initial state was prepared. Ideally the erasure check should leave the initial state alone, and this measurement sequence allows us to obtain the state fidelity for each cardinal state. This fidelity is post-selected on passing all the erasure checks, and on the end-of-line logical measurement not finding the cavities in $\ket{0,0}$. 

\begin{figure}
    \centering
    \includegraphics[width=1\linewidth]{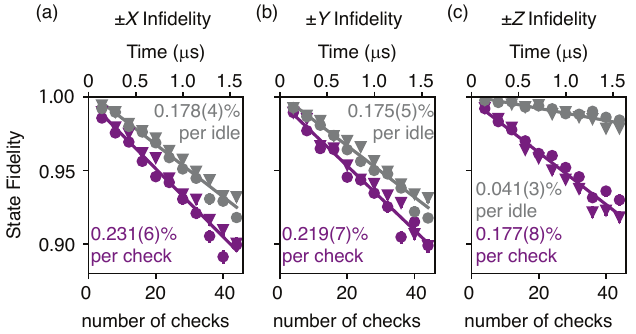}
    \caption{\textbf{Dual-rail state fidelities after repeated erasure checks.} (a-c) Purple markers show state fidelity of logical plus ($\ket{+X}$, $\ket{+Y}$ or $\ket{+Z}$) (dots) or logical minus ($\ket{-X}$, $\ket{-Y}$ or $\ket{-Z}$)(triangles) states after passing $n$ erasure checks. Gray markers show same fidelities when replacing each erasure check with a delay of the same duration. Lines show linear fits.}
    \label{fig:state_fid}
\end{figure}

The state fidelities for initial states $\ket{\pm X}_L$, $\ket{\pm Y}_L$ and $\ket{\pm Z}_L$ are shown in purple in Fig.~\ref{fig:state_fid}. The grey lines show the same fidelities when we instead idle for the duration of the beamsplitter and transmon pulses. Whereas the idling dual-rail state is strongly biased against having bit-flip errors (indicated by the high $\pm\ket{Z}_L$ state fidelity), the errors induced on the state during the erasure check are closer to a depolarizing channel. Just as with single-qubit gates, this removal of the bias is an inevitable consequence of applying the beamsplitter drive.

From the average state infidelities per erasure check, we extract the post-selected Pauli error rate via \cite{horodecki_general_1999}
\begin{equation}
    p_{\text{Pauli}} = p_x + p_y + p_z = \frac{3}{2}\left(1-\bar{F}\right).
\end{equation}
The factor of $3/2$ comes from the fact that only two out of the three types of errors ($X$, $Y$ or $Z$) will flip any of the cardinal states (e.g. the $\ket{+Z}$ state is unaffected by $Z$-type errors).

\begin{figure}
    \centering
    \includegraphics[width=1\linewidth]{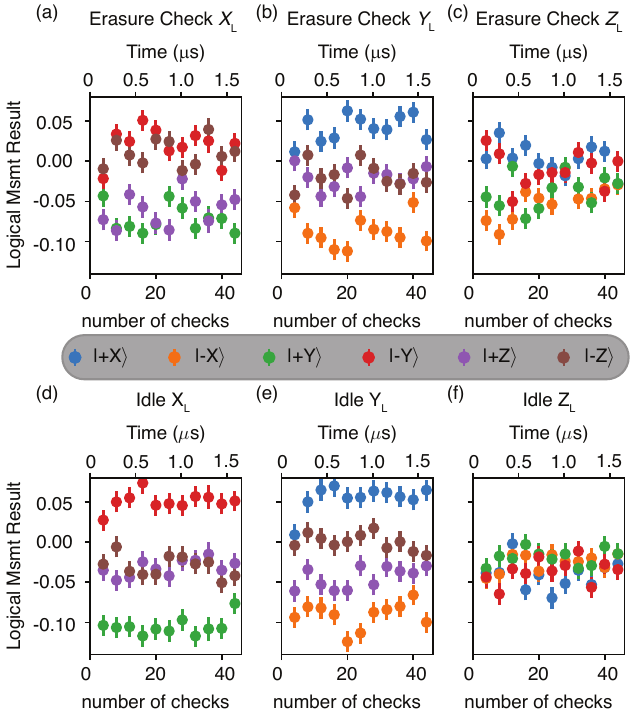}
    \caption{\textbf{Logical measurement results breakdown.} (a-c) Residual $X$, $Y$ and $Z$ logical measurement results for different initial states after passing $n$ mid-circuit erasure checks. (d-e) Same data when we instead idle for the duration of the beamsplitter and transmon pulses.}
    \label{fig:XYZ_logical}
\end{figure}

We also measure the expectation values of the logical operators perpendicular to the axis along which the initial state was prepared (Fig.~\ref{fig:XYZ_logical}). Ideally these values should be zero. In Fig.~\ref{fig:XYZ_logical}, the logical measurement results display small biases for both performing erasure check (top row) and when idling (bottom row). These biases are largest for $X_L$ and $Y_L$ measurements but they do not grow or noticeably oscillate in time. We speculate that this indicates that the sequence is not perfectly symmetric about the echo pulse.

\section{Error scaling for erasure check}
\label{app:error_scaling}
To help decide which parameters to use for the erasure check, we would like to determine how the erasure and Pauli error rates scale as we vary the decoherence rates of the system, the duration of the transmon pulse and the ratio of the beamsplitter amplitude to the dispersive shift. Here we do so with the aid of QuTiP \cite{johansson_qutip_2012} master equation simulations.

The key takeaways are that (a) increasing the pulse duration and commensurately reducing the transmon pulse amplitude reduces the probability of suffering transmon-induced erasure and Pauli errors, (b) increasing the beamsplitter amplitude reduces the probability of suffering transmon-induced Pauli errors, and (c) the likelihood of transmon-induced erasures and transmon-induced Pauli errors scale differently for transmon dephasing and transmon relaxation.

We can show this using a family of square transmon pulses that has consistent ideal performance in the absence of collapse operators while giving us the freedom to sweep the pulse duration. As a starting point, we center the pulse on the $N=0$ transition frequency, with a duration $T_\mathrm{p}$ chosen to put a notch in the spectrum at the central $N=1$ transition
\begin{equation}
    T_\mathrm{p} = \frac{2\pi}{|\chi|} \sqrt{4n^2-1},
    \label{eqn: tp_guess}
\end{equation}
where the integer $n$ indicates the number of detuned Rabi oscillations the transmon undergoes when the cavities are in $\ket{0,1}$ or $\ket{1,0}$. Meanwhile we can also choose the beamsplitter amplitude to ensure that the oscillator states approximately return to their starting positions at the end of $T_\mathrm{p}$:
\begin{equation}
    g_{\mathrm{bs}} = |\chi| \sqrt{\frac{m^2}{4n^2-1}-\frac{1}{4}},
    \label{eqn: gbs_guess}
\end{equation}
where the integer $m$ denotes the number of detuned beamsplitter oscillations the oscillator states undergo during the pulse duration. We then locally optimize the beamsplitter amplitude and pulse duration to suppress coherent errors from the above approximations. The pulses found are very close to these guessed parameters with negligible coherent errors.

\subsubsection*{Erasures due to transmon errors}
We would like to know how often (due to transmon errors) the transmon will end up in $\ket{e}$ given that the state was initialized in $\ket{\psi}_L\otimes\ket{g}$, averaged over dual-rail logical cardinal states $\ket{\psi}_L$. These are the false positives we see due to transmon dephasing and relaxation.

\begin{figure}
    \centering
    \includegraphics[width=0.75\linewidth]{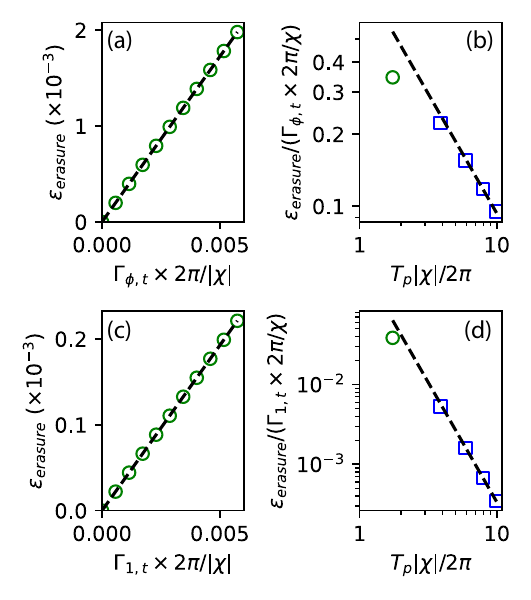}
    \caption{\textbf{Scaling of erasures arising from transmon errors.} (a) Simulated average Pauli error rate during erasure check as a function of the normalized transmon dephasing rate. Data is shown for a pulse duration $T_\mathrm{p}=1.7\times 2\pi/|\chi|$ and a beamsplitter rate $g_{\mathrm{bs}}=2.8|\chi|$ with the dashed line indicating a quadratic fit. (b) Linear component extracted from these quadratic fits (i.e. sensitivity of the erasure rate to changes in the normalized transmon dephasing rate) for pulses of varying duration. Black dashed line shows $1/x$ fit to last three points. The green point corresponds to the specific choice of pulse duration shown in the left column. (c)-(d) Same analysis for transmon relaxation-induced errors, where the fit in (d) is now to a $1/x^3$ curve.}
    \label{fig:erasures_scaling}
\end{figure}

We choose to set $m=5n$, corresponding to $|g_{\mathrm{bs}}/\chi|\approx2.5$, and sweep the value of $n$ from $1$ to $5$. For each choice of the pulse duration, we can sweep the transmon dephasing (relaxation) rate $\Gamma_{\phi,t}$ ($\Gamma_{1,t}$) and simulate how the erasure rate varies. We fit the resulting curve to a quadratic (see an example in Fig.~\ref{fig:erasures_scaling}a,c) and extract the linear component. The resulting slopes are then plotted as a function of pulse duration in Fig.~\ref{fig:erasures_scaling}b,d.

For transmon dephasing we find that
\begin{equation}
    \epsilon_{\text{erasure}} \propto \frac{\Gamma_{\phi,t}}{\chi}\left(T_\mathrm{p}\chi\right)^{-1},
\end{equation}
whereas for transmon relaxation errors we find good agreement to
\begin{equation}
    \epsilon_{\text{erasure}} \propto \frac{\Gamma_{1,t}}{\chi}\left( T_\mathrm{p}\chi\right)^{-3}.
\end{equation}
Provided that $g_{\mathrm{bs}}\gtrsim|\chi|$, we find that the beamsplitter rate has a negligible impact on the erasure rate. However, it will have a significant impact on the Pauli error rates.

\subsubsection*{Logical Pauli errors due to transmon errors}
\begin{figure}
    \centering
    \includegraphics[width=1\linewidth]{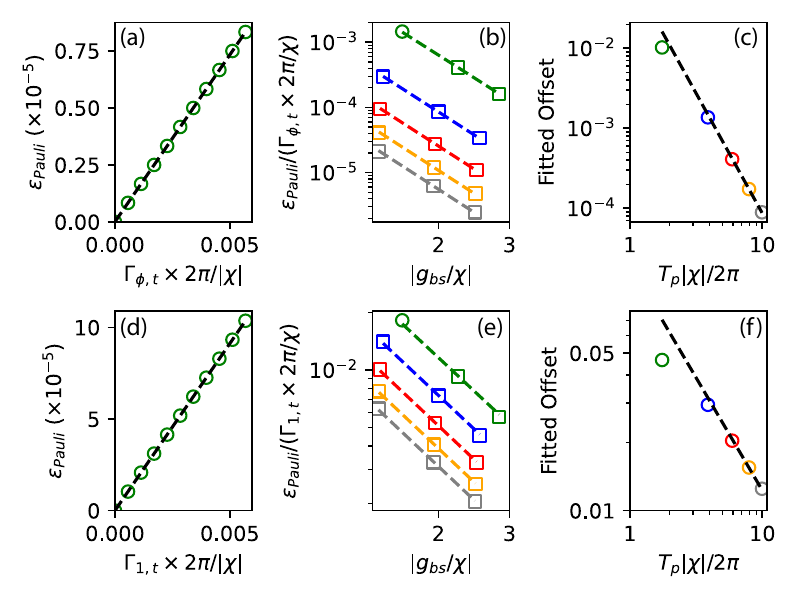}
    \caption{\textbf{Scaling of Pauli errors during erasure check arising from transmon errors.} (a) Simulated average Pauli error rate during erasure check as a function of transmon dephasing rate. Data is shown for a pulse duration $T_\mathrm{p}=1.7\times 2\pi/|\chi|$ and a beamsplitter rate $g_{\mathrm{bs}}=1.6|\chi|$ with dashed line indicating a quadratic fit. (b) Linear slope extracted from quadratic fits for pulses of varying $|g_{\mathrm{bs}}/\chi|$ and pulse duration (indicated by the colors). Within each pulse duration, the curves are fit to $A\times(g_{\mathrm{bs}}/\chi)^{-4}$ where the offset $A$ is the free parameter. (c) Fitted offsets $A$ as a function of pulse duration. Dashed black line shows a cubic fit. (d)-(f) Same analysis for transmon relaxation errors. The fit in (e) is now to $A\times(g_{\mathrm{bs}}/\chi)^{-2}$ and the fit in (f) is a linear fit.}
    \label{fig:pauli_scaling}
\end{figure}

We can perform the same analysis for Pauli errors, averaging the induced Pauli error rate over all dual-rail cardinal state as we sweep the transmon error rates (Fig~\ref{fig:pauli_scaling}a,d). To highlight the effect of the beamsplitter rate, we sweep both the pulse duration (via $n$) and $g_{\mathrm{bs}}$ (by using three different values set by $m=3n, 4n, 5n$). Fig.~\ref{fig:pauli_scaling}b,e show the slope of Pauli errors versus transmon error rate (obtained from the left panels) as a function of $|g_{\mathrm{bs}}/\chi|$. 

Increasing the beamsplitter amplitude quickly suppresses Pauli errors, with fits showing excellent agreement to a quartic suppression with $|g_{\mathrm{bs}}/\chi|$ for transmon dephasing, and a quadratic suppression for transmon decay.

The colored bands indicate pulses with the same pulse duration. Increasing the pulse duration also clearly has a beneficial impact on transmon-induced Pauli errors. To look at this scaling, we plot the offset of the previous polynomial fits versus $|g_{\mathrm{bs}}/\chi|$ and see how they vary as a function of $T_\mathrm{p}\chi$. These offsets are shown in Fig.~\ref{fig:pauli_scaling}c,f. The dashed lines show fits of the last three points to a cubic (transmon dephasing) or linear (transmon relaxation). While there is slight deviation from the fit at short pulse durations, we find that to good approximation:
\begin{equation}
    \epsilon_{\text{Pauli}} \propto \frac{\Gamma_{\phi,t}}{\chi}\left(\frac{g_{\mathrm{bs}}}{\chi}\right)^{-4}\left(T_\mathrm{p}\chi\right)^{-3} \propto \frac{\Gamma_{\phi,t}}{g_{\mathrm{bs}}^4T_\mathrm{p}^3},
\end{equation}
in the case of transmon dephasing, and 
\begin{equation}
    \epsilon_{\text{Pauli}} \propto \frac{\Gamma_{1,t}}{\chi}\left(\frac{g_{\mathrm{bs}}}{\chi}\right)^{-2}\left(T_\mathrm{p}\chi\right)^{-1} \propto \frac{\Gamma_{1,t}}{g_{\mathrm{bs}}^2T_\mathrm{p}},
\end{equation}
in the case of transmon relaxation.

\section{Comparison to joint-parity-based erasure check proposal}
\label{app:joint_parity}
The erasure check presented in this work differs from approach initially proposed in \cite{tsunoda_error-detectable_2023, teoh_dual-rail_2023}, which relies on measuring the parity of the combined photon number in the two oscillators. Here we will show how to interpret this joint-parity scheme in the context of the driven spectroscopy shown in this paper, and compare the expected performance of this scheme to the erasure check presented here.

\subsubsection*{Spectroscopic interpretation of joint-parity measurement}
\begin{figure}
    \centering
    \includegraphics[width=1\linewidth]{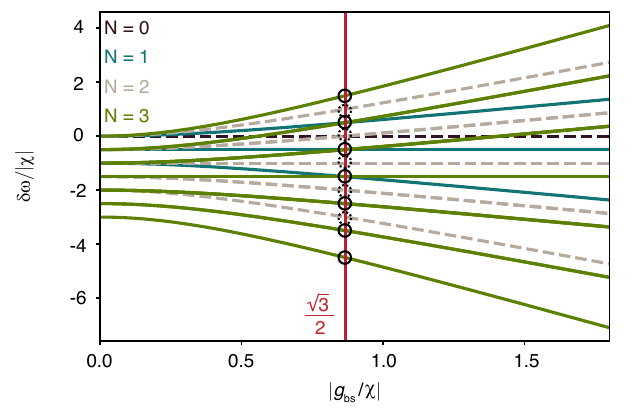}
    \caption{\textbf{Spectroscopic interpretation of joint-parity measurement.} The ancilla transition frequencies for a `symmetric' beamsplitter detuning $\Delta=\chi/2$ and total photon number in the two coupled oscillators of $N=0$ (black), $N=1$ (green), $N=2$ (beige, dashed) or $N=3$ (olive). The vertical red line highlights the beamsplitter amplitude $|g_{\mathrm{bs}}/\chi| = \sqrt{3}/2$, which is used for the joint parity measurement. At this value, all of the even $N$ transitions lie at a detuning equal to an even integer multiple of $\chi/2$ (dashed circles) and all of the odd $N$ transitions lie at odd integer multiples of $\chi/2$ (solid circles). The ancilla pulse ideally consists of two delta functions separated by $2\pi/\chi$, whose frequency spectrum has nodes at the solid circles and antinodes at the dashed circles.}   
    \label{fig:JP_interpretation}
\end{figure}

The joint-parity measurement consists of two $\pi/2$ pulses on the qubit, separated by a beamsplitter drive with amplitude $|g_{\mathrm{bs}}/\chi|=\sqrt{3}/2$ applied for a duration of $2\pi/\chi$, followed by a transmon readout. This scheme is analogous to the parity-map protocol for a single oscillator, where the addition of the beamsplitter drive allows the measurement qubit to accumulate an equal phase from photons in either oscillator.

When the beamsplitter amplitude is set to $|g_{\mathrm{bs}}/\chi|=\sqrt{3}/2$, the expression for the transition frequencies in Eq.~\ref{eqn:freqs} simplifies considerably:
\begin{equation}
    \omega_{\delta m | g_{\mathrm{bs}}/\chi=\sqrt{3}/2} = \left(\frac{N}{2}+\delta m\right) \chi, \quad \delta m = -N, \ldots, N.
\end{equation}
From this, we can see that for odd (even) $N$, all the transition frequencies lie at odd (even) integer multiples of $\chi/2$. This is higlighted in Fig.~\ref{fig:JP_interpretation}.

If we assume that the $\pi/2$-pulses are instantaneous, then the qubit pulse takes the form
\begin{equation}
    f(t) \propto \delta\left(t+\frac{\pi}{\chi}\right) + \delta\left(t-\frac{\pi}{\chi}\right),
\end{equation}
which has the Fourier transform:
\begin{equation}
    F(t) \propto \cos \left(\frac{\pi\omega}{\chi}\right).
\end{equation}

The frequency spectrum of the transmon pulse therefore has antinodes (nodes) at all of the transition frequencies with even (odd) joint-photon-number-parity. This provides an alternative interpretation of why the previously proposed scheme indeed maps the joint-photon-number parity of the two oscillators onto the transmon ancilla state.

\subsubsection*{Performance comparison}
The joint-parity measurement may be used with a two-level ($\ket{g}-\ket{e}$) or three-level ($\ket{g}-\ket{f}$) qubit. In both cases, transmon dephasing errors are fully error-detectable and do not contribute to the post-selected Pauli error rate. In the $\ket{g}-\ket{f}$ case, the $\ket{e}$ level is reserved as a flag state, allowing error detection of transmon relaxation errors as well. However, these flagged transmon errors do contribute to the overall erasure rate.

To compare the performance of the joint-parity scheme with the photon-number-selective scheme demonstrated in this work, we use master equation simulations considering only transmon relaxation and dephasing errors. In both cases we assume no limitation on pulse bandwidth, and use the system parameters in Table~\ref{tab:sys_params}, assuming $T_1^{fe} = T_1^{eg}/2$ and that $T_\phi^{gf}=T_\phi^{ge}/4$ (as would be the case for a harmonic oscillator). While typically $\chi_{gf} \approx 2\chi_{ge}$, this larger $\chi$ can only be accessed by commensurately increasing $g{bs}$. Supposing that $\chi$ can be optimized based on the available $g_{\text{bs}}$, we assume these values to be equal in each case. For the joint-photon-number-selective scheme we use a square pulse shape with $g_{\text{bs}}=1.04$ MHz and $T_p=1.699~\muup s$.

The results shown in Table~\ref{tab:sim_comparison} show that when using a $\ket{g}-\ket{e}$ transmon, the joint-photon-number measurement demonstrates fewer transmon-induced false positive erasures and fewer transmon-induced Pauli errors than the joint-parity measurement. 

The benefit of using a $\ket{g}-\ket{f}$ transmon to suppress Pauli errors due to transmon decoherence is substantially more pronounced for the joint-parity measurement since the scheme already protects against transmon dephasing errors. This results in a very low transmon-induced Pauli error rate. However it does so at the expense of a large additional erasure rate.

\begin{table}
\begin{tabular}{|c|c|c|c|c|}
\hline
 \multirow{2}{*}{} & \multicolumn{2}{|c|}{Joint-photon-number} & \multicolumn{2}{|c|}{Joint-parity}\\
 \cline{2-3} \cline{4-5}& $\ket{g}-\ket{e}$ & $\ket{g}-\ket{f}$ & $\ket{g}-\ket{e}$ & $\ket{g}-\ket{f}$ \\
 \hline
 $p_{\text{erasure}}$ & $0.61\%$ & $2.92\%$& $1.40\%$& $5.43\%$\\
 \hline
  $p_{\text{Pauli}}$ & $0.16\%$& $0.10\%$& $0.23\%$& $0.0048\%$\\
 \hline
\end{tabular}
\caption{\textbf{Simulated error rates during erasure check due to transmon relaxation and dephasing only.} The transmon-induced erasure probability $p_{\text{erasure}}$ and the transmon-induced Pauli error probability $p_{\text{Pauli}}$ (post-selected on not detecting an erasure) obtained from master equation simulations of the joint-photon-number measurement scheme demonstrated in this work, and the previously proposed joint-photon-number-parity measurement \cite{tsunoda_error-detectable_2023}. Both schemes are simulated using a $\ket{g}$-$\ket{e}$ ancilla and with a $\ket{g}$-$\ket{f}$ ancilla where the $\ket{e}$ level is reserved as a flag state.}
\label{tab:sim_comparison}
\end{table}

As discussed in \cite{tsunoda_error-detectable_2023}, in reality the finite bandwidth of the beamsplitter pulse during the joint-parity sequence would necessitate the use of optimal control pulses to ensure that the pulse completes within $2\pi/|\chi|$. Alternatively, without pulse shaping, delays can be added to make the total pulse duration $4\pi/|\chi|$. However this comes at the cost of twice as many transmon-induced errors and more idling errors.

When the error rates are dominated by transmon decoherence (as opposed to intrinsic errors), the greater suppression of Pauli errors offered by the joint-parity scheme ensures a greater bias towards erasure errors during the check. This could therefore enable a higher error-correcting threshold (closer to $7\%$ than the approximately $4\%$ in this work). However, this improvement comes with a very large increase in the overall error rate. Since the performance of an erasure code depends on the ratio of the error rate to the threshold, the joint-photon-number-selective scheme is preferred.

One caveat is that the joint-photon-number-parity measurement also catches heating to dual-rail states with $N=2$, unlike the version of the check presented in this work. These uncaught events are rare but are especially damaging for error correction. However, one could adapt the photon-number-selective scheme by adding a 2-photon-selective transmon pulse to prevent this.

\setlength{\tabcolsep}{12pt}
\begin{table*}
\begin{tabular}{|c|c|c|c|c|c|c|}
\hline
\multirow{2}{*}{Parameter} & \multirow{2}{*}{Unit}& \multicolumn{2}{|c|}{Alice} & \multirow{2}{*}{SNAIL} & \multicolumn{2}{|c|}{Bob} \\
\cline{3-4} \cline{6-7}& 
& Transmon & Cavity & &Cavity & Transmon \\
\hline
$\omega$ & $\text{GHz}$ & $4.783\,720(1)$ & $2.973\,421(1)$ & $5.192\,429(1)$ & $6.926\,368(2)$ & $5.402\,311(3)$ \\
\hline
$T_{1}$ & $\muup$s & $123.4\pm0.9$ & $347\pm2$ & $69\pm2$ & $108.5\pm0.6$ & $42.4\pm0.4$ \\
\hline
$T_{2R}$ & $\muup$s & $31.6\pm0.9$ & $441\pm23$ & $3.0\pm0.2$ & $165\pm7$ & $33.1\pm0.9$ \\
\hline
$T_{2E}$ & $\muup$s & $40.5\pm0.7$ & - & $14.4\pm0.6$ & - & $63.2\pm0.8$ \\
\hline
$n_{th}$ & $\%$ & $0.23\pm0.06$ & $3.82\pm0.10$ & $4.1\pm0.6$ & $0.53\pm0.03$ & $0.52\pm0.07$\\
\hline
$\chi$ & $\text{MHz}$ & \multicolumn{2}{|c|}{$0.7773(4)$} & & \multicolumn{2}{|c|}{$1.0660(9)$} \\ 
\hline
\end{tabular}
\caption{\textbf{Measured system parameters.}}
\label{tab:sys_params}
\end{table*}

\section{Predicting erasure check performance with longer coherences}
\label{app:expected_performance}
We simulate the estimated the performance improvement with longer transmon coherences ($T_1=T_{\phi}=200~\muup$s) and cavity lifetimes ($T_{1,A}=T_{1,B}=1000~\muup$s). We use the same dispersive shift $\chi$ and readout duration $\tau_{RO}$ used in this experiment. We choose a square pulse shape (without ramps) with pulse duration $T_p=1.699~\muup$s and beamsplitter rate $g_{\text{bs}}/2\pi=1.038~$MHz, the closest ideal pulse parameters to the experimentally optimized parameters we use. 

For each dual-rail input state, we initialize the state, wait $\tau_{RO}$ (to capture the effect of photon loss during readout), and then perform the pulse. From this final state we extract the probability of the cavities being in $\ket{0,0}$ (the intrinsic erasure rate, $p_{\text{erasure}}^{\text{intrinsic}} = 0.334\%$) and the probability of suffering a Pauli error given that the transmon was in $\ket{g}$ (the induced Pauli error rate, $p_{\text{Pauli}}^{\text{induced}} = 0.035\%$). To obtain the false positive rate, we run the same simulation without the cavity collapse operators, and measure the probability of the transmon ending in $\ket{e}$ ($p_{\text{FP}} = 0.164\%$).

An effect not included in these simulations that will start to become important is measurement-induced dephasing on the dual-rail qubit due to photons in the readout resonator. This dephasing rate is given by 
\begin{equation}
    \Gamma_{\phi} = \frac{\bar{n}\kappa\chi^2}{\kappa^2+\chi^2},
    \label{eq:msmt_induced_dephasing}
\end{equation}
where $\bar{n}$ is the average photon number in the readout resonator, $\kappa = 2\pi\times 1.77~$MHz is its linewidth, and $\chi$ is the dispersive shift between the resonator and Bob's cavity \cite{Gambetta-dephasing_2006}. We estimate $\chi$ from the anharmonicity of Bob's transmon $\alpha=-2\pi\times 185~$MHz, its dispersive shift on Bob's cavity $\chi_{\text{ct}}=-2\pi\times 1.066~$MHz and its dispersive shift on Bob's readout resonator $\chi_{\text{tr}}=-2\pi\times 0.86~$MHz, as
\begin{equation}
\chi \approx \frac{\chi_{\text{tr}}\chi_{\text{ct}}}{\alpha} \approx 2.5~\text{kHz}.
\end{equation}
We then take $\bar{n}\approx10$~ photons for $1~\muup$s during the readout. Putting these values into Eq.~\ref{eq:msmt_induced_dephasing}, and using $p_{\text{Pauli}} = \Gamma_{\phi}t/2$, we obtain an approximate Pauli error rate per erasure check due to measurement-induced dephasing of $p_{\text{Pauli}} \approx 0.01\%$.

\section{Requirements on $|g_{\mathrm{bs}}/\chi|$ at larger photon numbers}
\label{app:large_N}
As discussed in the Methods, the state of the two coupled oscillators with $N$ photons between them is analogous to the state of a single spin with $S=N/2$. In this picture, we saw that the matrix element for a transition from the eigenstate of $\hat{H}_g$ (the Hamiltonian when the qubit is in $\ket{g}$) with magnetic quantum number $m_g$ to the  $m_e$ eigenstate of $\hat{H}_e$ (the Hamiltonian when the qubit is in $\ket{e}$) can be expressed in terms of the small Wigner $d$-matrix, $|d_{m_g, m_e}^{N/2}(\delta\theta)|$. Furthermore, we saw that for the choice of beamsplitter detuning $\Delta=\chi/2$, all the transition frequencies for the same total photon number $N$ and change in magnetic quantum number $\delta m \equiv m_e - m_g$ become degenerate, allowing us to see joint-photon number-splitting.

\begin{figure}
    \centering
    \includegraphics[width=1\linewidth]{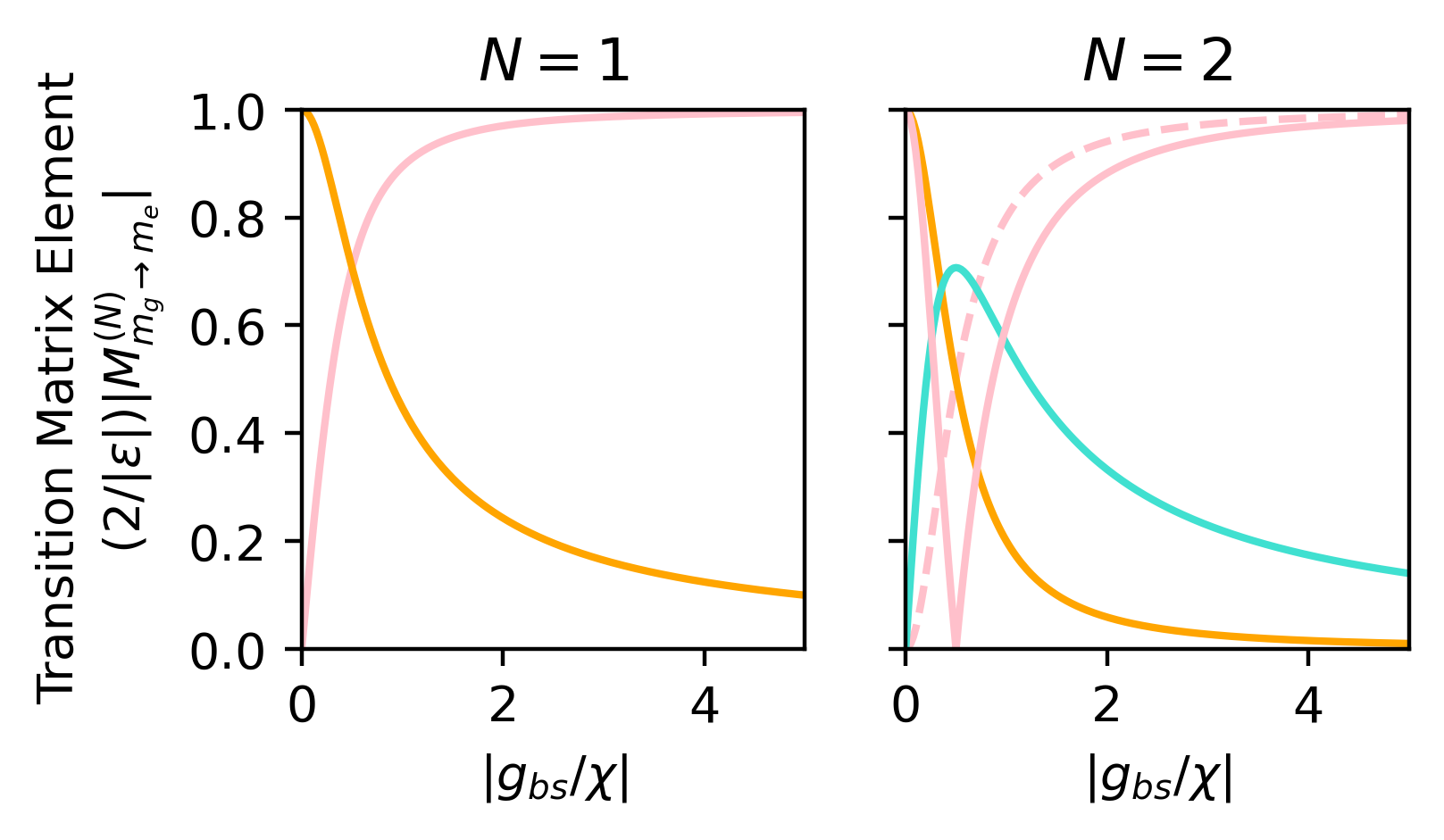}
    \caption{\textbf{Theoretical transition matrix elements.} Predicted transition matrix elements (normalized for the amplitude of the spectroscopy tone) for a) $N=1$ and b) $N=2$ total photons in the two oscillators, when the beamsplitter detuning is symmetric ($\Delta=\chi/2$). Colors distinguish transitions that are degenerate in frequency with $\delta m=0$ (pink), $\delta m=\pm1$ (orange) and $\delta m=\pm 2$ (turquoise). For $N=2$ and $\delta m=0$, the matrix element for the transition $0\rightarrow0$ (solid) differs from those for $\pm1\rightarrow\pm1$ (dashed).}
    \label{fig:matrix_elements}
\end{figure}

However, by looking at the elements of the Wigner $d$-matrix, we find that these transitions with the same frequency do \textit{not} always have the same transition matrix element. The transition matrix elements for $N=1$ and $N=2$ are shown in Fig.~\ref{fig:matrix_elements}. Since the $d$-matrix obeys the symmetry $|\delta_{i,j}^{N/2}(\beta)| = |\delta_{j,i}^{N/2}(\beta)|$, we find that $|M_{\frac{1}{2}\rightarrow\frac{1}{2}}| = |M_{-\frac{1}{2}\rightarrow-\frac{1}{2}}|$ and so all the degenerate transitions for $N=1$ also have the same matrix element. By contrast, in the 2-photon manifold we find that the transition matrix element from $m_g=0$ to $m_e=0$ is given by
\begin{equation}
    \left|M_{0\rightarrow 0}^{(N=2)}\right| = \frac{|\epsilon|}{2}\left|d_{0,0}^{1}\left(\delta\theta\right)\right| = \frac{|\epsilon|}{2}\left|\cos\left(\delta\theta\right)\right|,
\end{equation}
whereas the matrix element from $m_g=\pm1$ to $m_e=\pm 1$ is
\begin{equation}
    \left|M_{\pm1\rightarrow \pm1}^{(N=2)}\right| = \frac{|\epsilon|}{2}\left|d_{\pm1,\pm1}^{1}\left(\delta\theta\right)\right| = \frac{|\epsilon|}{2}\frac{\left|1+\cos\delta\theta\right|}{2},
\end{equation}
where $\delta\theta = 2\arctan (\chi/2g_{\text{bs}})$.

The implication is that joint-photon-number-selective qubit pulses for $N\geq2$ will proceed at different rates for different oscillator states within the $N$-photon manifold, thereby introducing coherent errors. However, at large beamsplitter rates, the difference in rates becomes suppressed as $(\chi/g_{\mathrm{bs}})^2$.

\section{System parameters}
\label{app:system_params}

Table \ref{tab:sys_params} lists the coherences and thermal populations of the modes in the system, as well as the dispersive coupling strength between each cavity and its ancilla transmon.  

The SNAIL parameters are measured at the external flux bias used to obtain the data ($\Phi_{\text{ext}} = 0.334\Phi_0$). Its low $T_{2R}$ is due to low-frequency flux noise away from its flux sweet spot, but introducing an echo increases its $T_{2E}$ coherence substantially.

As mentioned in the Methods, this device was previously used in Chapman et al. \cite{chapman_high--off-ratio_2023}. One change that was made to the device was an increase the coupling to Bob's readout resonator, increasing its linewidth from $\kappa=2\pi\times (0.89 \pm .05)$ MHz to $\kappa=2\pi\times (1.77 \pm .04)$ MHz. This allowed us to reduce the duration of the readout and therefore reduce idling errors.

As discussed in Chapman et al. \cite{chapman_high--off-ratio_2023}, the oscillator coherence times in this system are lower than is typical for $\lambda/4$ superconducting stub cavities. The bare cavities, without any chips inserted, were measured to have lifetimes of $2.3$~ms and $460~\muup$s, respectively. These lifetimes therefore do not explain the measured coherences. Meanwhile, finite element simulations including Purcell loss out of the system ports and conductor loss due to the magnetic coil suggest that these effects should not limit the oscillator lifetimes to less than $2.66$~ms and $810~\muup$s. We often find large cooldown-to-cooldown variations in $T_1$ when changes are made inside the package (e.g. changing a coupling pin length, or re-inserting a chip), even when these changes are not expected to be related. For example, the slight increase in \textit{Bob's} readout pin length described earlier coincided with a decrease in \textit{Alice's} cavity $T_1$ from $485\pm14~\muup$s to $347\pm2~\muup$s. Meanwhile, on another cooldown with all of the chips inserted, Bob's cavity $T_1$ was measured to be $265\pm27~\muup$s. This difficult-to-predict behavior could point to unexpected package modes as the culprit.

\end{document}